\documentclass[aps,prx,twocolumn,superscriptaddress,nofootinbib,longbibliography,floatfix]{revtex4-2}

\usepackage{amsmath,amssymb,bm}
\usepackage{graphicx}
\usepackage{microtype}
\usepackage{hyperref}

\newcommand{\Tr}{\operatorname{Tr}}
\newcommand{\Cov}{\operatorname{Cov}}
\newcommand{\Var}{\operatorname{Var}}
\newcommand{\cH}{\mathsf H}
\newcommand{\cF}{\mathsf F}
\newcommand{\cB}{\mathsf B}
\newcommand{\cD}{\mathsf D}
\newcommand{\cP}{\mathsf P}
\newcommand{\cC}{\mathsf C}
\newcommand{\PhiB}{\overline{\Phi}}
\newcommand{\rhoB}{\overline{\rho}}

\begin{document}

\title{Task-Resolved Fisher Spectroscopy for Quantum Reservoir Computing}
\author{Yang Peng}
\affiliation{Department of Physics and Astronomy, California State University, Northridge, California 91330, USA}
\affiliation{Institute of Quantum Information and Matter and Department of Physics, California Institute of Technology, Pasadena, California 91125, USA}

\begin{abstract}
Quantum reservoir computing (QRC) uses fixed quantum dynamics to encode a time series and trains only a classical readout, but a benchmark capacity alone does not reveal whether task information is lost in the reservoir, the measurement, feature compression, or finite sampling. We introduce \emph{task-resolved Fisher spectroscopy}, in which prediction targets define orthonormal score coordinates on the stationary distribution of input histories. Reweighting labeled histories along these scores generates an exactly affine family of reservoir states and measurement outcomes. In the same coordinates, we obtain a many-body Fisher-information hierarchy relating the state quantum Fisher information, the Fisher information of the complete measurement record, and moment matrices retained through many-body order $r$. The quadratic form of each moment matrix is exactly the stationary capacity of the optimal linear readout, while a finite-measurement extension predicts the approach from one-shot records to ideal expectation-value features. The measurement-level quantities require only stationary labeled records and measured outcome strings, not an input model or quantum-state tomography. In a five-spin open reservoir, interactions route fourth-order temporal information into higher-body correlations, so low-order compression can incur orders-of-magnitude sampling overhead even when a conventional parity target remains accessible. For correlated inputs and outputs, record-defined task scores predict held-out capacities, the required feature order, and measurement-budget dependence; optimizing the local measurement axis recovers otherwise hidden task information. The framework connects benchmark performance to experimentally actionable diagnoses of reservoir encoding, measurement choice, classical representation, and shot allocation.
\end{abstract}

\maketitle

\section{Introduction}
\label{sec:introduction}

Time-series prediction asks for a present or future quantity to be inferred from a finite or fading history of observations. Reservoir computing (RC) addresses this problem by replacing a fully trained recurrent network with a fixed dynamical system whose transient response provides a high-dimensional representation of the recent input history \cite{Jaeger2001,Maass2002,Dambre2012}. Only a final readout is trained, usually by linear regression. This separation is especially attractive for physical computing because the complicated part of the processor can be supplied by natural dynamics rather than by optimizing every internal coupling \cite{Appeltant2011,Brunner2013,Torrejon2017}.

Quantum reservoir computing (QRC) applies the same architecture to driven quantum systems. In the standard setting a classical input sequence controls a quantum channel or Hamiltonian, the evolving quantum state carries fading memory of earlier inputs, selected observables are measured, and only the coefficients of the final classical readout are fitted \cite{Fujii2017,Nakajima2019,Ghosh2019,Chen2020,Mujal2021}. The internal Hamiltonian, couplings, dissipation, and measurement protocol are normally fixed during task training. Many-body spin, continuous-variable, circuit, and hybrid reservoirs have been studied theoretically, with performance linked to dynamical regimes, dissipation, coherence, noise, and the number and type of measured observables \cite{MartinezPena2021,Nokkala2021,Govia2021,Suzuki2022,Dudas2023,Kubota2023,Tran2023,Sannia2024,MartinezPenaIPC2023,Palacios2024}. Experimental progress includes analog superconducting reservoirs for microwave-signal processing, digital quantum processors operating beyond a single circuit coherence time, circuit-QED reservoirs, and correlated-spin reservoirs performing nonlinear temporal benchmarks and real-world forecasting \cite{Senanian2024,Hu2024,Carles2026,Hou2025}.

Recent theory has also moved beyond reporting a single benchmark error. Information-processing capacity (IPC) resolves which temporal functions can be reconstructed by a linear readout \cite{Dambre2012,MartinezPenaIPC2023}; finite sampling and measurement noise have been incorporated into physical QRC descriptions \cite{Khan2021,Polloreno2023,Palacios2024,Du2026}; and input dependence, fading memory, and faithful encoding of distinct histories have been analyzed at the level of finite-dimensional reservoir filters \cite{MartinezPenaOrtega2023,Kobayashi2024,MartinezPena2025}. Very recent work has introduced a delay-space quantum Fisher information matrix for finite-shot memory \cite{WangQiu2026}, process-tensor and Holevo diagnostics for storage, scrambling, local accessibility, and loss \cite{KeenanZambrini2026}, state-space or delay-resolved decompositions of reservoir capacity \cite{Nokkala2026,AbdallaRontani2026}, and thermodynamic accounts of temporal information processing \cite{DingQiu2026,Cenedese2026}.

It is well established that, for a parameterized family of quantum states, quantum Fisher information provides a state-level information metric and bounds the Fisher information attainable by measurements, while a specified measurement defines the corresponding classical Fisher matrix \cite{BraunsteinCaves1994,Petz1996}. Covariance--response constructions similarly quantify information accessible through a restricted set of measured observables \cite{Gessner2018,Gessner2020}, and standard reservoir-computing capacity measures quantify how well temporal target functions can be reconstructed by a linear readout \cite{Dambre2012}. These ingredients address state sensitivity, measurement accessibility, and prediction from complementary viewpoints. An experimentally useful QRC characterization should connect them for the same prediction task and the same stationary operating distribution.

This gap is especially important experimentally. A quantum state may retain information about the input history that is not exposed by a chosen measurement, while a complete measurement record may contain task-relevant information that is lost when the classical interface keeps only low-order observables such as $\langle X_i\rangle$ or $\langle X_iX_j\rangle$. Moreover, a reservoir that appears excellent when exact expectation values are supplied by a simulation may become statistically inefficient when those observables must be estimated from finite measurement records. For a specified prediction problem, one would therefore like to determine how much information about the actual target is present in the reservoir, what fraction is exposed by the measurement, at what many-body order it becomes accessible, and how the answer changes with the available measurement budget.

We address these questions with \emph{task-resolved Fisher spectroscopy}. The prediction targets themselves define the Fisher coordinates. For one output, the centered and normalized target gives a single task score. For several correlated outputs $y_{1t},y_{2t},\ldots$, Gram--Schmidt orthonormalization with respect to the stationary operating distribution gives orthonormal score directions spanning the same target subspace. Controlled reweighting of labeled input histories along these scores generates an affine statistical family of reservoir states and measurement outcomes. The measured response to each score is its feature--score covariance, so the resulting moment matrix gives the standard linear-readout capacity of each original target and of any linear combination within their span. The same score coordinates also define state-level quantum Fisher information and the classical Fisher information of the complete measurement record. In this way the prediction task is lifted from the final classical readout to a common information geometry of the reservoir state, the chosen measurement, and progressively restricted many-body features. All measurement-level ingredients can be estimated from stationary labeled records and measured output bit strings. For independent unbiased binary inputs, Walsh modes form an analytic orthonormal score family \cite{Walsh1923}. Standard benchmark targets such as PC2 and PC4 correspond to particular modes, while more general prediction targets can involve arbitrary linear combinations of Walsh modes.

The remainder of the paper is organized as follows. Section~\ref{sec:prelim} introduces the QRC prediction setting, linear-readout capacity, measurement-dependent Fisher information, and the many-body features used in the readout. Section~\ref{sec:spectroscopy} develops task-resolved Fisher spectroscopy for single- and multi-output prediction, derives the state--measurement--feature information hierarchy, relates it to standard prediction capacity, and gives an experimental construction together with the finite-measurement-budget extension. The Gram--Schmidt construction of multiple task scores and supporting derivations are given in the Appendices. Section~\ref{sec:numerics} uses the same dissipative interacting spin reservoir in two complementary regimes: an independent unbiased binary source, where Walsh scores resolve an entire temporal subspace, and a correlated binary source, where the task coordinates are constructed directly from the operating distribution. The conclusion is presented in Section~\ref{sec:conclusion}.

\section{Quantum-reservoir prediction and measurement preliminaries}
\label{sec:prelim}

\subsection{Fixed quantum dynamics and a trained linear readout}

\begin{figure}[t]
\includegraphics[width=0.498\textwidth]{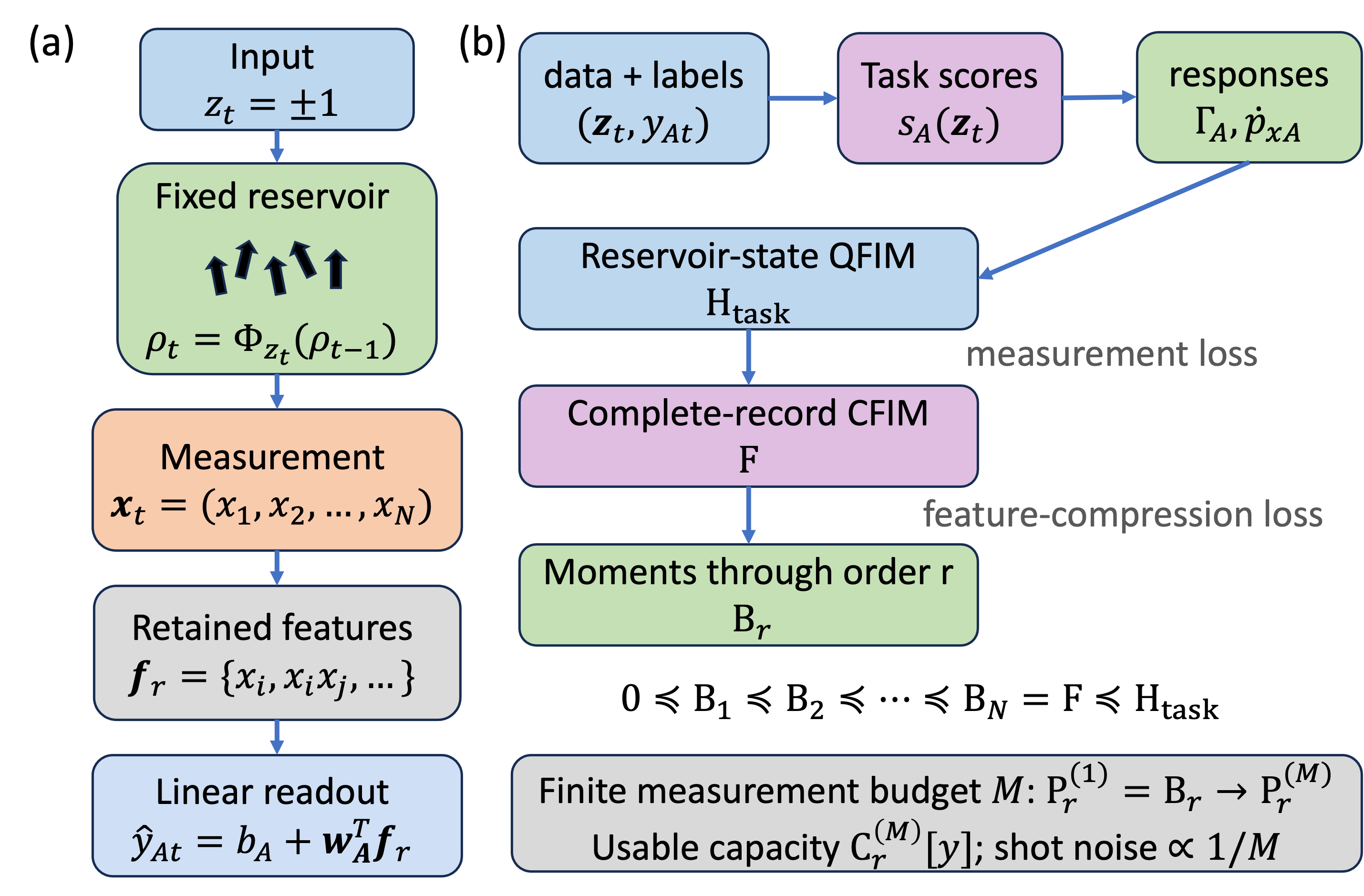}
\caption{
(a) \textbf{Quantum-reservoir workflow.} A stationary input history drives a fixed open quantum reservoir. Measurement produces a bit string, products through many-body order $r$ form the retained feature vector, and only the final linear readout is trained. (b) \textbf{Task-resolved Fisher spectroscopy.} The target labels define orthonormal task scores $s_A$. Their response tangents determine the reservoir-state QFIM $\cH_{\rm task}$ and the complete-record CFIM $\cF$; compressing the same measurement record to moments through order $r$ gives $\cB_r$. The hierarchy separates measurement loss from feature-compression loss. The lower box shows the finite-measurement extension: $\cP_r^{(1)}=\cB_r$, while $\cP_r^{(M)}$ determines the usable capacity $\mathcal C_r^{(M)}[y]$ for $M$ measurements of each history-conditioned reservoir state.}

\label{fig:qrcworkflow}
\end{figure}

The architecture is summarized in Fig.~\ref{fig:qrcworkflow}. We consider a binary input sequence
$z_t\in\{-1,+1\}$. During one reservoir update, the current symbol selects one of two completely positive trace-preserving (CPTP) maps, $\Phi_\pm$ for the input $z=\pm 1$, so that the density matrix $\rho_t$ at each time $t$ is obtained according to the following rule:
\begin{equation}
\rho_t=\Phi_{z_t}(\rho_{t-1}).
\label{eq:qrcupdate}
\end{equation}
The state $\rho_t$ therefore contains a dynamical encoding of the recent input history. The microscopic realization of $\Phi_\pm$ is not needed at this stage; a specific driven open-spin realization is introduced in Sec.~\ref{sec:numerics}.

A chosen measurement of the reservoir produces a time-dependent feature vector $\bm f_r$ containing measured observables up to many-body order $r$, i.e., observables represented by operators acting on at most $r$ spins.
In an ideal numerical simulation these features may be exact expectation values, whereas in an experiment they are estimated from measurement outcomes. A standard reservoir readout predicts a target $y_t$ according to
\begin{equation}
\widehat y_t=b+\bm w^T\bm f_r ,
\label{eq:linreadout}
\end{equation}
where $\widehat y_t$ denotes the estimated value of the target $y_t$.
Only the intercept $b$ and the weights $\bm w$ are fitted to training data; the reservoir dynamics remain fixed during this regression. Thus a linear readout does not imply a linear dependence on the input history: memory and nonlinearity are generated by the reservoir, while only the final map from reservoir features to the target is linear. For simultaneous prediction of several variables $y_{At}$, $A=1,\ldots,K$, one may use one linear readout $\widehat y_{At}=b_A+\bm w_A^T\bm f_r$ for each output.

After an initial washout period, we assume a stationary statistical regime: under the stationary input process, the joint statistics of the input history, reservoir state, measurement outcomes, and target are invariant under a shift of the time index. The state along an individual trajectory generally continues to fluctuate with the input sequence; stationarity refers to the ensemble statistics rather than to a time-independent state.

Throughout the paper, $\mathbb E[\cdot]$ denotes an ensemble average over the stationary input process and the corresponding reservoir measurement outcomes, unless an explicitly named operating or reweighted ensemble is indicated by a subscript. We use
\begin{equation}
\Cov(A,B)
=
\mathbb E\!\left[
(A-\mathbb E[A])(B-\mathbb E[B])
\right]
\label{eq:covdef}
\end{equation}
for the covariance, with the natural matrix-valued extension for vector quantities.

We center the target and feature vector so that
$\mathbb E[y]=0$ and $\mathbb E[\bm f]=0$. For the general readout in Eq.~\eqref{eq:linreadout}, minimizing the mean-squared error with respect to the intercept gives
$b_{\rm LS}=\mathbb E[y]-\bm w^T\mathbb E[\bm f]$, and therefore $b_{\rm LS}=0$ for these centered variables. Define
\begin{equation}
\cC_f=\Cov(\bm f,\bm f),
\qquad
\bm d_y=\Cov(\bm f,y).
\label{eq:predcov}
\end{equation}
The remaining least-squares minimization gives
\begin{equation}
\bm w_{\rm LS}=\cC_f^+\bm d_y,
\label{eq:lsweights}
\end{equation}
where $+$ denotes the Moore--Penrose pseudoinverse. The corresponding minimum mean-squared error is
\begin{equation}
\mathrm{MSE}_{\min}
=
\Var(y)-\bm d_y^T\cC_f^+\bm d_y .
\end{equation}
Following the standard information-processing-capacity terminology of reservoir computing \cite{Dambre2012,MartinezPenaIPC2023}, the capacity for reconstructing the centered target $y$ from the chosen features is
\begin{equation}
\mathcal C[y|\bm f]
=
1-\frac{\mathrm{MSE}_{\min}}{\Var(y)}
=
\frac{\bm d_y^T\cC_f^+\bm d_y}{\Var(y)} .
\label{eq:explainedvariance}
\end{equation}
It is the fraction of the target variance reconstructed by the best linear readout allowed by the stationary input--output statistics. This is an infinite-data statistical quantity; estimates obtained from finite training and test trajectories fluctuate around it.

For the binary inputs considered below, two standard parity-check (PC) benchmarks are
\begin{equation}
Y_{\rm PC2}(t)=z_tz_{t-1},
\qquad
Y_{\rm PC4}(t)=z_tz_{t-1}z_{t-2}z_{t-3}.
\label{eq:pcbenchmarks}
\end{equation}
Because $z_t=\pm1$, both are parity variables: they are $+1$ when an even number of the multiplied symbols are negative and $-1$ when that number is odd. PC2 probes a second-order temporal dependence over two symbols, whereas PC4 probes a fourth-order dependence over four symbols. A delayed PC4 target,
$z_{t-1}z_{t-2}z_{t-3}z_{t-4}$, has the same nonlinear order but requires one additional time step of memory. These tasks are used below as controlled diagnostics rather than as models of an unknown forecasting target.

\subsection{Measurement-dependent Fisher information}

Prediction depends not only on information encoded in the reservoir state but also on what is exposed by the chosen measurement. To distinguish these levels, consider a smooth family of quantum states $\rho(\bm\xi)$ depending on parameters $\bm\xi$, and write
$\partial_A\equiv\partial/\partial\xi_A$.
The symmetric logarithmic derivatives $L_A$ satisfy
\begin{equation}
\partial_A\rho
=
\frac12(\rho L_A+L_A\rho),
\end{equation}
and define the quantum Fisher information matrix (QFIM)
\begin{equation}
(\cH)_{AB}
=
\frac12\Tr[\rho\{L_A,L_B\}] .
\label{eq:qfim_general}
\end{equation}
The QFIM characterizes local parameter distinguishability at the state level.

For a specified POVM $\{M_x\}$, the outcome probabilities
$p_x(\bm\xi)=\Tr[M_x\rho(\bm\xi)]$ define the corresponding classical Fisher information matrix (CFIM),
\begin{equation}
(\cF)_{AB}
=
\sum_x
\frac{\partial_Ap_x\,\partial_Bp_x}{p_x}.
\label{eq:cfi_general}
\end{equation}
For any fixed measurement, $\cF\preceq\cH$,
where $A\preceq B$ means that $B-A$ is positive semidefinite, or equivalently that
$\bm v^TA\bm v\le\bm v^TB\bm v$ for every vector $\bm v$. This inequality expresses the fact that a specified measurement can expose only part of the parameter sensitivity available in the quantum state \cite{BraunsteinCaves1994}. In Sec.~\ref{sec:spectroscopy}, the generic parameters $\bm\xi$ will be replaced by task-score coordinates.

\subsection{Many-body features of a fixed measurement}

In our concrete setup discussed below, the main readout is a simultaneous local-$X$ measurement on $N$ spins. One shot produces a bit string
\begin{equation}
\bm x=(x_1,\ldots,x_N),
\qquad x_i=\pm1.
\end{equation}
For every nonempty subset $S\subseteq\{1,\ldots,N\}$, define the correlation feature
\begin{equation}
f_{S}(\bm x)=\prod_{i\in S}x_i ,
\label{eq:measuredfeature}
\end{equation}
where the time dependence is implicit through $\bm x=\bm x_t$.

Collecting all products with $1\le |S|\le r$ gives the feature vector $\bm f_r(\bm x)$. Its components include $x_i$, $x_ix_j$, and progressively higher-order products, whose averages give
$\langle X_i\rangle$, $\langle X_iX_j\rangle$, and correlations up to many-body order $r$.

Because all local $X_i$ commute, a single local-$X$ measurement setting already supplies the complete bit string from which all these products can be formed. Increasing $r$ therefore changes only the classical representation of the same measurement record, not the quantum measurement setting.

\section{Task-resolved Fisher spectroscopy}
\label{sec:spectroscopy}
In this section, we introduce task-score coordinates that allow the state-level Fisher information, the information in the complete measurement record, and the information retained by truncated many-body features to be compared directly.

\subsection{Task scores and operating-distribution ensembles}

Let $\bm z_t=(z_t,z_{t-1},\ldots,z_{t-L+1})$ denote the most recent $L$ binary inputs at prediction time $t$, and let $P_{\rm ref}(\bm z)$ be their stationary distribution in the operating time series. The source need not be independent or uniformly distributed, and $P_{\rm ref}$ need not be known analytically; a long stationary record provides samples from it. We use $\mathbb E_{\rm ref}$, $\Cov_{\rm ref}$, and $\Var_{\rm ref}$ for the expectation, covariance, and variance defined in Sec.~\ref{sec:prelim}, evaluated under this operating distribution. We assume first that the prediction variables are determined by the same recent history, $y_{At}=g_A(\bm z_t)$, $A=1,\ldots,K$. The single-output problem is recovered by setting $K=1$. The reservoir itself may retain memory beyond $L$.

For a given window $\bm z$, let $\bar\rho(\bm z)=\mathbb E[\rho_t\mid\bm z_t=\bm z]$ denote the reservoir state averaged over any earlier context compatible with that window under the stationary operating process. The corresponding reference state is $\rho_{\rm ref}=\sum_{\bm z}P_{\rm ref}(\bm z)\bar\rho(\bm z)$. Thus the task window need not determine the reservoir state uniquely.

For several outputs, define centered targets $\widetilde y_A(\bm z)=g_A(\bm z)-\mathbb E_{\rm ref}[y_A]$ and use the reference inner product $\langle u,v\rangle_{\rm ref}=\mathbb E_{\rm ref}[uv]$. Applying Gram--Schmidt to the linearly independent centered targets produces score functions $s_A(\bm z)$ spanning the same target subspace and satisfying $\mathbb E_{\rm ref}[s_A]=0$ and $\mathbb E_{\rm ref}[s_As_B]=\delta_{AB}$. The explicit recursion, including its direct sample-based implementation, is given in Appendix~\ref{app:gramschmidt}. For a single target $y$, this reduces to the normalized score $s_y=(y-\mu_y)/\sigma_y$, with $\mu_y=\mathbb E_{\rm ref}[y]$ and $\sigma_y^2=\Var_{\rm ref}(y)$.

The orthonormal scores define the controlled family
\begin{equation}
P_{\bm\theta}(\bm z)
=P_{\rm ref}(\bm z)\left[1+\sum_{A=1}^K\theta_A s_A(\bm z)\right],
\label{eq:scoreensemble}
\end{equation}
with $\bm\theta$ restricted so that all probabilities remain nonnegative. Normalization follows from $\mathbb E_{\rm ref}[s_A]=0$. Because the history space is finite for a binary length-$L$ input window, the scores are bounded and a nonzero neighborhood of $\bm\theta=\bm0$ is always admissible.

The ensemble-averaged reservoir state is exactly affine,
\begin{equation}
\rho(\bm\theta)=\rho_{\rm ref}+\sum_A\theta_A\Gamma_A,
\qquad
\Gamma_A=\sum_{\bm z}P_{\rm ref}(\bm z)s_A(\bm z)\bar\rho(\bm z).
\label{eq:affine}
\end{equation}
No expansion in the physical input amplitude is involved. Experimentally, Eq.~\eqref{eq:scoreensemble} is implemented by importance weighting stationary labeled records, as described in Sec.~\ref{sec:experiment}.

\subsection{Task-resolved Fisher hierarchy and many-body accessibility}

The score parameters $\bm\theta$ play the role of the generic parameters $\bm\xi$ introduced in Sec.~\ref{sec:prelim}. We denote the QFIM of Eq.~\eqref{eq:affine} at $\bm\theta=\bm0$ by $\cH_{\rm task}$. For the fixed local-$X$ measurement, let $p_x(\bm\theta)$ be the probability of the complete measured bit string. The affine state family implies $p_x(\bm\theta)=p_x+\sum_A\theta_A\dot p_{xA}$, with $\dot p_{xA}=\left.\partial p_x/\partial\theta_A\right|_{\bm0}$, and hence
\begin{equation}
(\cF)_{AB}=\sum_x\frac{\dot p_{xA}\dot p_{xB}}{p_x},
\qquad
\cF\preceq\cH_{\rm task}.
\label{eq:cfi}
\end{equation}
Both matrices act in the same space of task-score directions.

We next retain only the measured features $\bm f_r$ containing local-$X$ products through many-body order $r$. The covariance matrix introduced generically in Sec.~\ref{sec:prelim} becomes $\cC_r=\Cov_{\rm ref}(\bm f_r,\bm f_r)$. The response of these features to the score ensemble is
\begin{equation}
(\cD_r)_{\alpha A}
=\left.\frac{\partial}{\partial\theta_A}\mathbb E_{\bm\theta}[f_\alpha]\right|_{\bm0}
=\Cov_{\rm ref}(f_\alpha,s_A).
\label{eq:Drdef}
\end{equation}
The second equality is the standard response--covariance identity for a score perturbation and is derived in Appendix~\ref{app:response_covariance}.

From the same covariance--response construction used in multiparameter moment methods \cite{Gessner2018,Gessner2020}, define
\begin{equation}
\cB_r=\cD_r^T\cC_r^+\cD_r.
\label{eq:Br}
\end{equation}
The nested measured feature spaces, together with the monotonicity of Fisher information under measurement, give
\begin{equation}
0\preceq\cB_1\preceq\cB_2\preceq\cdots\preceq\cB_N=\cF\preceq\cH_{\rm task}.
\label{eq:mainhier}
\end{equation}
The moment part of the hierarchy is proved in Appendix~\ref{app:projector}; the final inequality follows from the usual monotonicity of Fisher information under measurement. Thus $\cH_{\rm task}-\cF$ quantifies task-resolved information not exposed by the chosen measurement, whereas $\cF-\cB_r$ quantifies information already present in the complete local-$X$ record but discarded when only correlations through order $r$ are retained.

\subsection{Prediction capacity in the task-score basis}

The score construction makes the connection to prediction direct. Let $\bm s=(s_1,\ldots,s_K)^T$ denote the orthonormal task scores and let $\cD_r$ be the response matrix in Eq.~\eqref{eq:Drdef}. Every centered original target has known coordinates in this score basis,
\begin{equation}
y_A-\mathbb E_{\rm ref}[y_A]=\sum_{B=1}^K c_B^{(A)}s_B,
\label{eq:targetscorecoords}
\end{equation}
where the coefficients $c_B^{(A)}=\mathbb E_{\rm ref}[(y_A-\mathbb E_{\rm ref}[y_A])s_B]$ are obtained directly during the Gram--Schmidt construction. Orthonormality gives $\Var_{\rm ref}(y_A)=(\bm c^{(A)})^T\bm c^{(A)}$, while Eq.~\eqref{eq:Drdef} gives the feature--target covariance vector $\bm d_{y_A}=\cD_r\bm c^{(A)}$. Substitution into Eq.~\eqref{eq:explainedvariance} yields
\begin{equation}
\mathcal C_r[y_A]
=\frac{(\bm c^{(A)})^T\cB_r\bm c^{(A)}}{(\bm c^{(A)})^T\bm c^{(A)}}.
\label{eq:taskcapacitymulti}
\end{equation}
Thus the same matrix $\cB_r$ gives the linear-readout capacity of every output variable in a multi-output prediction problem. More generally, any centered linear combination of the outputs corresponds to another direction in the same score subspace and is evaluated by the same quadratic form.

For a single target, $y-\mu_y=\sigma_y s_y$ and Eq.~\eqref{eq:taskcapacitymulti} reduces to
\begin{equation}
\mathcal C_r[y]=\bm D_{r,y}^T\cC_r^+\bm D_{r,y},
\qquad
\bm D_{r,y}=\frac{\Cov_{\rm ref}(\bm f_r,y)}{\sigma_y}.
\label{eq:taskcapacityscore}
\end{equation}
Hence the task-directed moment information is exactly the stationary linear-readout capacity for the same operating distribution and measured feature set. The Fisher construction lifts this familiar covariance quantity to the reservoir state and the complete measurement record, so losses due to measurement choice and low-order classical compression can be separated.

For the single normalized task direction, let $\cF[y]$ and $\cH[y]$ denote the corresponding scalar classical and quantum Fisher informations. The hierarchy~\eqref{eq:mainhier} gives
\begin{equation}
0\le\mathcal C_1[y]\le\cdots\le\mathcal C_N[y]=\cF[y]\le\cH[y].
\label{eq:taskhierarchy}
\end{equation}

\subsection{Walsh modes for independent unbiased binary inputs}

For the independent unbiased binary inputs used in our numerical study, the score construction has a closed analytic form, with $P_{\rm ref}(\bm z)=2^{-L}$. For each nonempty subset $A\subseteq\{1,\ldots,L\}$ define the Walsh modes~\cite{Walsh1923} $\chi_A(\bm z_t)=\prod_{k\in A}(\bm z_t)_k$, where $(\bm z_t)_k=z_{t-k+1}$. These functions already satisfy $\mathbb E_{\rm ref}[\chi_A]=0$ and $\mathbb E_{\rm ref}[\chi_A\chi_B]=\delta_{AB}$, so the score modes are simply $s_A=\chi_A$. Equation~\eqref{eq:scoreensemble} reduces to
\begin{equation}
P_{\bm\theta}(\bm z)=2^{-L}\left[1+\sum_A\theta_A\chi_A(\bm z)\right].
\label{eq:walshdist}
\end{equation}
A parity target such as PC2 or PC4 is therefore already a normalized task score.

For this i.i.d. reference process, the average channel $\PhiB=(\Phi_++\Phi_-)/2$ has at least one stationary density operator $\rhoB$ with $\PhiB(\rhoB)=\rhoB$. Writing $\Delta=(\Phi_+-\Phi_-)/2$ gives $\Phi_z=\PhiB+z\Delta$. Expanding a finite history in these two channel components yields the explicit Walsh response operators $\Gamma_A$ and reproduces Eq.~\eqref{eq:affine} at arbitrary separation between the two physical input levels. The construction is derived in Appendix~\ref{app:walsh}. The resulting Walsh score family is used below to resolve an entire temporal subspace and to benchmark the many-body information hierarchy.

\subsection{Generalized temporal modes and sampling overhead}

For a multidirectional score family, $\cB_r$ and $\cF$ can be compared through
\begin{equation}
\cB_r\bm v_j=\lambda_j^{(r)}\cF\bm v_j,
\qquad
\bm v_i^T\cF\bm v_j=\delta_{ij},
\label{eq:geneig}
\end{equation}
restricted to the support of $\cF$, i.e. the orthogonal complement of its kernel. Directions in $\ker\cF$ produce no Fisher response even in the complete local-$X$ record and cannot be normalized with respect to $\cF$. The hierarchy implies $0\le\lambda_j^{(r)}\le1$; $\lambda_j^{(r)}$ is the fraction of the full measured Fisher information retained after the bit string is reduced to correlations through order $r$.

The same fraction gives a finite-sampling interpretation. Appendix~\ref{app:finiteshot} shows that resolving a deliberately imposed score perturbation along $\bm v_j$ with order-$r$ features requires $\kappa_j^{(r)}=1/\lambda_j^{(r)}$ times as many independent spectroscopy samples as using the complete measured bit string to reach the same local estimator variance. For a single task score this reduces to the ratio $\cF[y]/\mathcal C_r[y]$. This spectroscopy overhead concerns the statistical efficiency of a restricted representation, complementing finite-shot analyses of moment estimation and QRC readout \cite{Khan2021,Du2026}; the separate effect of repeatedly measuring the same history-conditioned reservoir state is considered next.

\subsection{Finite measurement budget for prediction}

In an experimental QRC, the feature vector supplied to the classical readout may itself be estimated from repeated measurements. Let $\mathcal H_t$ denote the complete input history up to time $t$. For a fixed history, define the conditional feature mean
\begin{equation}
\bm\mu_r(\mathcal H_t)
=
\mathbb E\!\left[
\bm f_r(\bm x_t)
\mid
\mathcal H_t
\right],
\label{eq:historyconditionalmean}
\end{equation}
where $\bm x_t$ is the stochastic measurement bit string obtained at time $t$. The conditional expectation averages only over measurement outcomes while holding the input history fixed.

If the same history-conditioned state is prepared independently $M$ times, with measurement outcomes $\bm x_t^{(1)},\ldots,\bm x_t^{(M)}$, the feature vector supplied to the readout is
\begin{align}
\widetilde{\bm f}_{r,M}(\mathcal H_t)
&=
\frac{1}{M}\sum_{m=1}^{M}
\bm f_r\!\left(\bm x_t^{(m)}\right)
\nonumber\\
&=
\bm\mu_r(\mathcal H_t)+\bm\epsilon_M,
\qquad
\mathbb E[\bm\epsilon_M\mid\mathcal H_t]=0.
\label{eq:replayedfeature}
\end{align}
For conditionally independent repetitions, the law of total covariance gives
\begin{equation}
\cC_r(M)
=
\cC_r^{\rm dyn}
+
\frac{1}{M}\cC_r^{\rm shot},
\label{eq:CM}
\end{equation}
where
$\cC_r^{\rm dyn}
=
\Cov_{\rm ref}[
\bm\mu_r(\mathcal H_t),
\bm\mu_r(\mathcal H_t)
]$
describes the variation of the conditional feature means across input histories, while
$\cC_r^{\rm shot}
=
\mathbb E_{\rm ref}[
\Cov(
\bm f_r(\bm x_t)
\mid
\mathcal H_t
)
]$
is the average one-shot measurement covariance at fixed history.
The average defining $\cC_r^{\rm dyn}$ is taken over the full stationary input process, including inputs earlier than the finite target window whenever the reservoir retains longer memory.

Each task score $s_A$ is determined by the input history. Since $\mathbb E[\bm\epsilon_M\mid\mathcal H_t]=0$, averaging repeated measurements does not change its covariance with the measured features:
\begin{equation}
\Cov_{\rm ref}
\!\left(
\widetilde{\bm f}_{r,M},
s_A
\right)
=
\Cov_{\rm ref}
\!\left(
\bm\mu_r(\mathcal H_t),
s_A
\right)
=
(\cD_r)_{:A}.
\label{eq:DindependentM}
\end{equation}
Thus increasing $M$ suppresses measurement fluctuations without changing the task-dependent response matrix. We define the measurement-budget-dependent moment-information matrix
\begin{equation}
\cP_r^{(M)}
=
\cD_r^T
\cC_r(M)^+
\cD_r .
\label{eq:PM}
\end{equation}
For an original prediction target $y_A$ with coordinates $\bm c^{(A)}$ in the orthonormal task-score basis, its capacity at measurement budget $M$ is
\begin{equation}
\mathcal C_r^{(M)}[y_A]
=
\frac{
(\bm c^{(A)})^T
\cP_r^{(M)}
\bm c^{(A)}
}{
(\bm c^{(A)})^T\bm c^{(A)}
}.
\label{eq:finitebudgetcapacity}
\end{equation}
For a single target, $\bm c=\sigma_y\bm e_1$, so this expression reduces to the corresponding diagonal element of $\cP_r^{(M)}$. The one-shot and exact-expectation limits are
\begin{equation}
\cP_r^{(1)}=\cB_r,
\qquad
\cP_r^{(\infty)}
=
\cD_r^T
(\cC_r^{\rm dyn})^+
\cD_r .
\label{eq:Plimits}
\end{equation}
Hence the measurement budget changes the feature noise and therefore the usable prediction capacity, while leaving the task-dependent signal encoded in $\cD_r$ unchanged.

\subsection{Experimental construction from a stationary labeled time series}
\label{sec:experiment}

Task-resolved Fisher spectroscopy can be constructed directly from a stationary labeled operating record. Suppose a stationary calibration record provides $N_s$ sampled segments. We write the $n$th length-$L$ segment ending at prediction time $t$ as
\begin{equation}
\bm z_t^{(n)}=(z_t^{(n)},z_{t-1}^{(n)},\ldots,z_{t-L+1}^{(n)}),
\qquad n=1,\ldots,N_s,
\label{eq:sampledsegments}
\end{equation}
with measured output bit string $\bm x_t^{(n)}$ and target labels $y_{At}^{(n)}$, $A=1,\ldots,K$. The index $t$ labels time within a segment, while $n$ labels different sampled segments or realizations. Stationarity makes their marginal statistics time independent, while ergodicity justifies replacing ensemble averages by averages over a sufficiently long record. Correlations between overlapping windows must be retained when estimating statistical uncertainties.

From the target labels, center the $K$ output variables and apply the empirical Gram--Schmidt procedure of Appendix~\ref{app:gramschmidt}. This gives score values $\widehat s_A^{(n)}$ satisfying $N_s^{-1}\sum_n\widehat s_A^{(n)}=0$ and $N_s^{-1}\sum_n\widehat s_A^{(n)}\widehat s_B^{(n)}=\delta_{AB}$ for the retained linearly independent target directions. The same procedure records the coefficients $\widehat{\bm c}^{(A)}$ that express each original centered output in the score basis.

For score direction $A$, assign segment $n$ the weight $w_{n,A}(\theta)=1+\theta\widehat s_A^{(n)}$ and form the weighted feature mean
\begin{equation}
\widehat{\bm\mu}_{r,A}(\theta)
=\frac{\sum_n w_{n,A}(\theta)\bm f_r(\bm x_t^{(n)})}{\sum_n w_{n,A}(\theta)}.
\label{eq:weightedmean}
\end{equation}
Because the empirical scores are centered, $\sum_nw_{n,A}(\theta)=N_s$. Symmetric biases $\pm\epsilon$ therefore give the $A$th response column,
\begin{equation}
(\widehat\cD_r)_{:A}
=\frac{\widehat{\bm\mu}_{r,A}(+\epsilon)-\widehat{\bm\mu}_{r,A}(-\epsilon)}{2\epsilon}.
\label{eq:experimentalD}
\end{equation}
For these linear importance weights, the finite difference is exactly independent of $\epsilon$ within the positivity range and equals the empirical covariance between $\bm f_r$ and the score $\widehat s_A$. Repeating the same post-processing for all retained task scores reconstructs $\widehat\cD_r$.

The unweighted records give
$\widehat{\cC}_r=\widehat{\Cov}_{\rm ref}(\bm f_r,\bm f_r)$
and hence
$\widehat{\cB}_r=\widehat{\cD}_r^{\,T}\widehat{\cC}_r^+\widehat{\cD}_r$.
The capacity of any original output $y_A$ follows from
Eq.~\eqref{eq:taskcapacitymulti} using its empirical score coordinates
$\widehat{\bm c}^{(A)}$.
Because each local-$X$ measurement returns the complete bit string
$\bm x=(x_1,\ldots,x_N)$, the same records can be post-processed to form
feature vectors $\bm f_r$ containing all products
$f_S(\bm x)=\prod_{i\in S}x_i$ with $1\le |S|\le r$.
Thus $r=1$ retains only single-spin outcomes, $r=2$ retains single- and
two-spin products, and increasing $r$ progressively includes higher-order
correlations up to $r=N$.
The entire many-body hierarchy can therefore be reconstructed from the same
measurement records by changing only the classical post-processing.
Weighted histograms of the complete bit strings give the corresponding
multiparameter classical Fisher matrix $\cF$.
Thus the task scores, their Gram--Schmidt orthonormalization, the feature
covariances and responses, the task capacities, and the full-record Fisher
matrix can all be obtained from the same labeled experimental record.
Only the state-level QFIM generally requires additional quantum-state
characterization.
\section{Numerical demonstration in an open quantum spin reservoir}
\label{sec:numerics}
We apply the task-resolved framework to one open-spin model under two stationary input processes. For an independent unbiased binary stream, Walsh modes provide an analytic task basis and expose the geometry of an entire temporal sector. For a correlated binary stream with correlated outputs, the task coordinates are instead learned from the operating distribution. The two examples test, respectively, the basis-resolved and data-defined forms of the construction.

\subsection{Physical model and numerical procedure}

We evaluate the theory in an interacting spin reservoir with Hamiltonian
\begin{equation}
H(a)=\sum_{i=1}^{N}\left(h_i\sigma_i^z+a\sigma_i^x\right)
+\sum_{i<j}J_{ij}\sigma_i^x\sigma_j^x.
\label{eq:ham}
\end{equation}
The reservoir couples to an equilibrium thermal bath at inverse temperature $\beta$ through the fixed system operator $S=\sum_i\sigma_i^y$ and evolves under a Lindblad master equation \cite{Gorini1976,Lindblad1976}. Appendix~\ref{app:lindblad_jumps} derives the corresponding many-body jump operators from the Bohr-frequency decomposition of $S$, following the construction used in Ref.~\cite{Cenedese2026}.

The scalar $a$ is the externally controlled transverse-field amplitude carrying the binary input, $a_t=a_0+\delta a\,z_t$, with $z_t=\pm1$. Here $a_0$ specifies the operating point, while $2\delta a$ is the separation between the two physical input levels. The task-resolved Fisher construction treats the corresponding channels $\Phi_\pm=\Phi_{a_0\pm\delta a}$ at their finite separation and does not require $\delta a$ to be small. We use $a_0=1.5$ and $\delta a=0.5$, giving $a_-=1$ and $a_+=2$. Each value of $a_t$ is held fixed for one update interval, $\Delta t=1$.

We use $N=5$, homogeneous $h_i=h=1$, inverse temperature $\beta=10$, and $\gamma\Delta t=1$. The chosen input levels are both of order the local field scale $h$ and define a representative working window rather than an optimization of the physical drive.

Because $\partial H/\partial a=\sum_i\sigma_i^x$, the observables $X_i=\sigma_i^x$ are local generalized forces conjugate to the injected control amplitude. For a switch of the control, $\Delta a\,X_i$ is a local contribution to the switching-energy increment. Products of the $X_i$ therefore resolve correlations among the input-conjugate generalized-force channels.

For each input value, we diagonalize Eq.~\eqref{eq:ham}, construct the global jump operators, and propagate the state for one interval $\Delta t$. The stationary state of the average channel is obtained from a trace-constrained linear solve. Task-response operators then determine the QFIM and the local-$X$ probability responses, while moment projections are evaluated by singular-value decomposition to avoid unstable inversion of nearly singular covariance matrices.

\subsection{Independent unbiased inputs and Walsh-resolved temporal geometry}
\label{sec:walsh_numerics}

We first specialize to the i.i.d. unbiased source, with equal probabilities $\Pr(z_t=\pm1)=1/2$. In this regime each Walsh mode is an exact normalized score. This basis reveals how interactions reorganize temporal information across measured many-body orders and how that geometry controls the capacity and sampling cost of particular parity targets such as PC2 and PC4.

\subsubsection{Disorder ensembles versus operation of a single reservoir}

The couplings are drawn independently as $J_{ij}\sim U[-J_s,J_s]$. We use 100 paired coupling patterns at each $J_s/h$, and rescale the same normalized patterns across interaction strengths. The disorder ensemble tests whether the observed trends are typical; it is not part of the operation of a deployed QRC, which uses one fixed reservoir and trains only its classical readout.

The ensemble calculation resolves the $q=4$ sector for $L=6$ delays, containing $\binom64=15$ fourth-order temporal modes. For every realization we compute the $15\times15$ QFIM, the complete local-$X$ CFIM, and the moment matrices through five-body order. We restrict the main comparison to $J_s/h\le10$, where the measured fourth-order CFIM remains full rank, and do not use the rank-deficient stronger-coupling regime for the headline comparisons.
\subsubsection{Interaction-induced correlation depth}

\begin{figure}[t]
\includegraphics[width=\columnwidth]{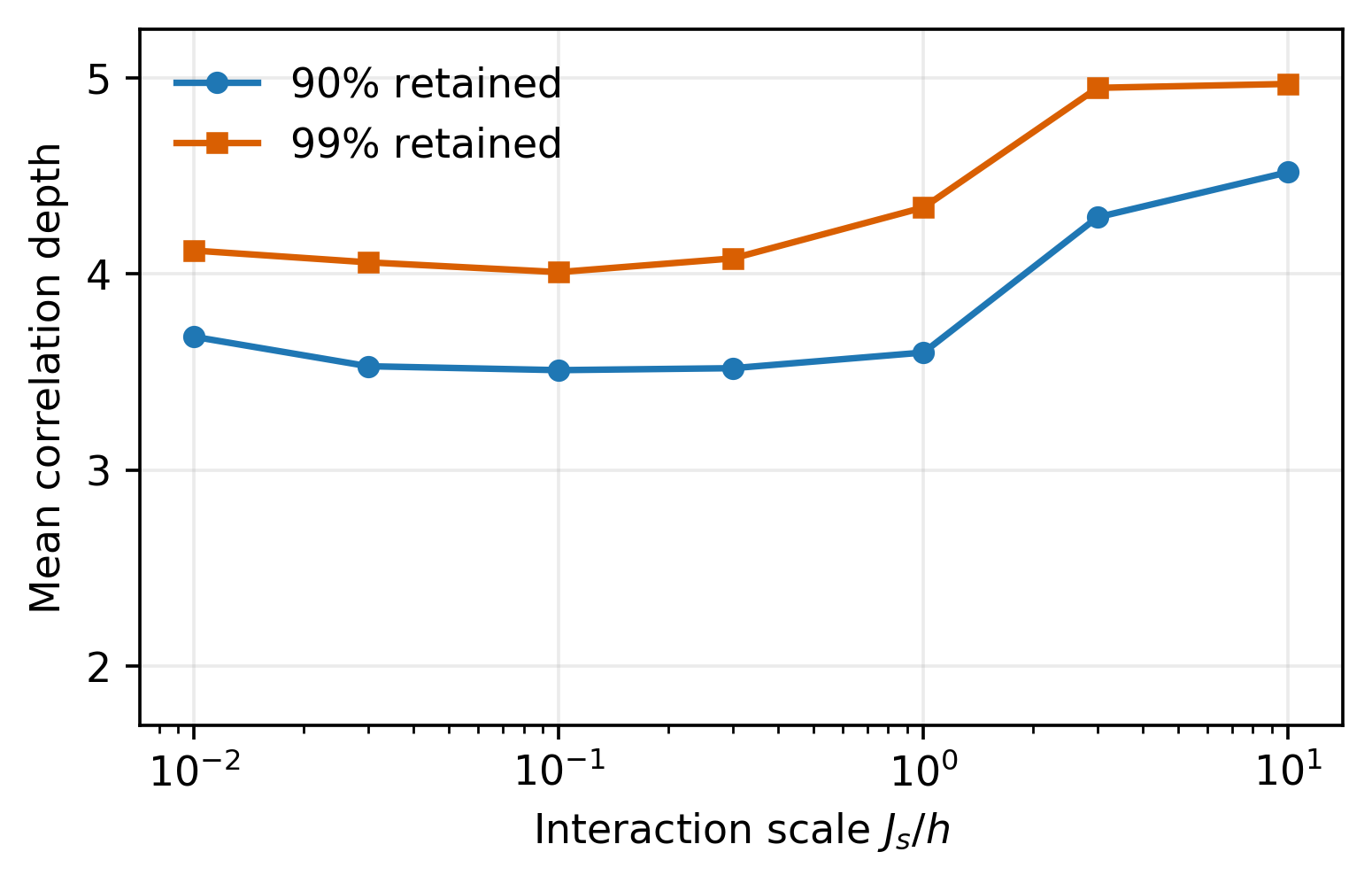}
\caption{\textbf{Interaction-dependent correlation depth of fourth-order temporal information.} Correlation depth is the minimum measured many-body order required for a specified fraction of complete local-$X$ Fisher information. The plotted ensemble contains 100 paired coupling realizations at each $J_s/h$. Headline comparisons are restricted to the full-rank regime $J_s/h\le10$.}
\label{fig:depth}
\end{figure}

For each generalized temporal mode, we define its correlation depth as the smallest measured many-body order $r$ that retains a specified fraction of the complete local-$X$ Fisher information. Figure~\ref{fig:depth} shows that strong interactions increase the mean correlation depth for both retention thresholds 90\% and 99\%. Thus interactions move some temporal information into higher-body correlations of the measured bit string.

\subsubsection{Low-order compression can impose orders-of-magnitude shot overhead}

\begin{figure}[t]
\includegraphics[width=\columnwidth]{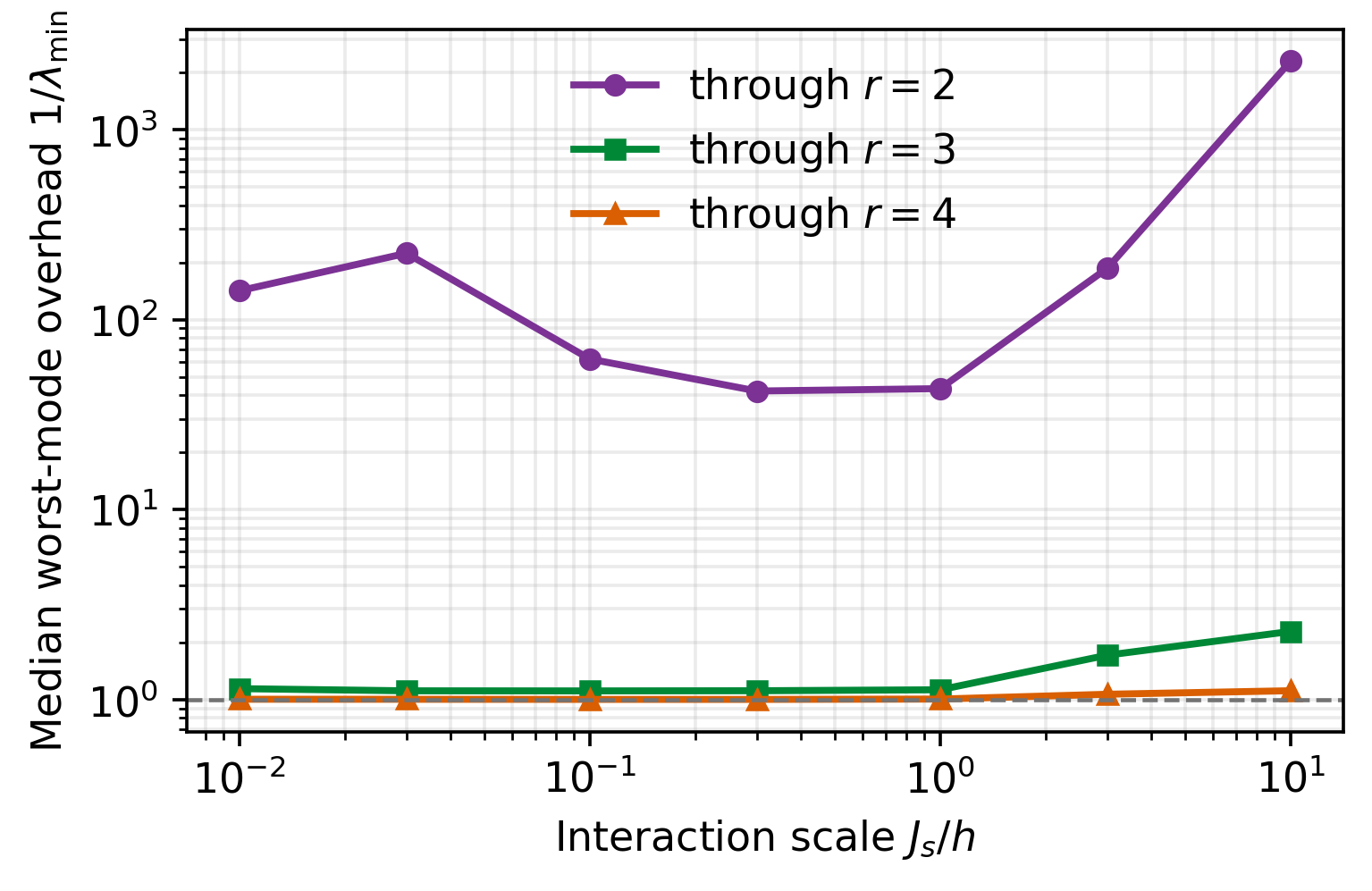}
\caption{\textbf{Sampling penalty from low-order data compression.} Median worst-mode sampling overhead $\kappa=1/\lambda_{\min}$ when the complete local-$X$ bit string is compressed to correlation products through order $r$. All correlations are extracted from the same local-$X$ bit strings; the overhead arises from discarding higher-order classical statistics, not from using additional measurement settings.}
\label{fig:overhead}
\end{figure}

Figure~\ref{fig:overhead} gives the corresponding sampling cost. At strong coupling, compressing the bit string to moments through two-body order produces a penalty of several orders of magnitude for the least accessible temporal mode. Including three-body moments removes most of this penalty, and four-body moments approach the efficiency of the complete record. Thus, information present in the measured bit strings can be discarded by low-order classical compression.

\subsubsection{A single parity benchmark can miss the bottleneck}

\begin{figure}[t]
\includegraphics[width=\columnwidth]{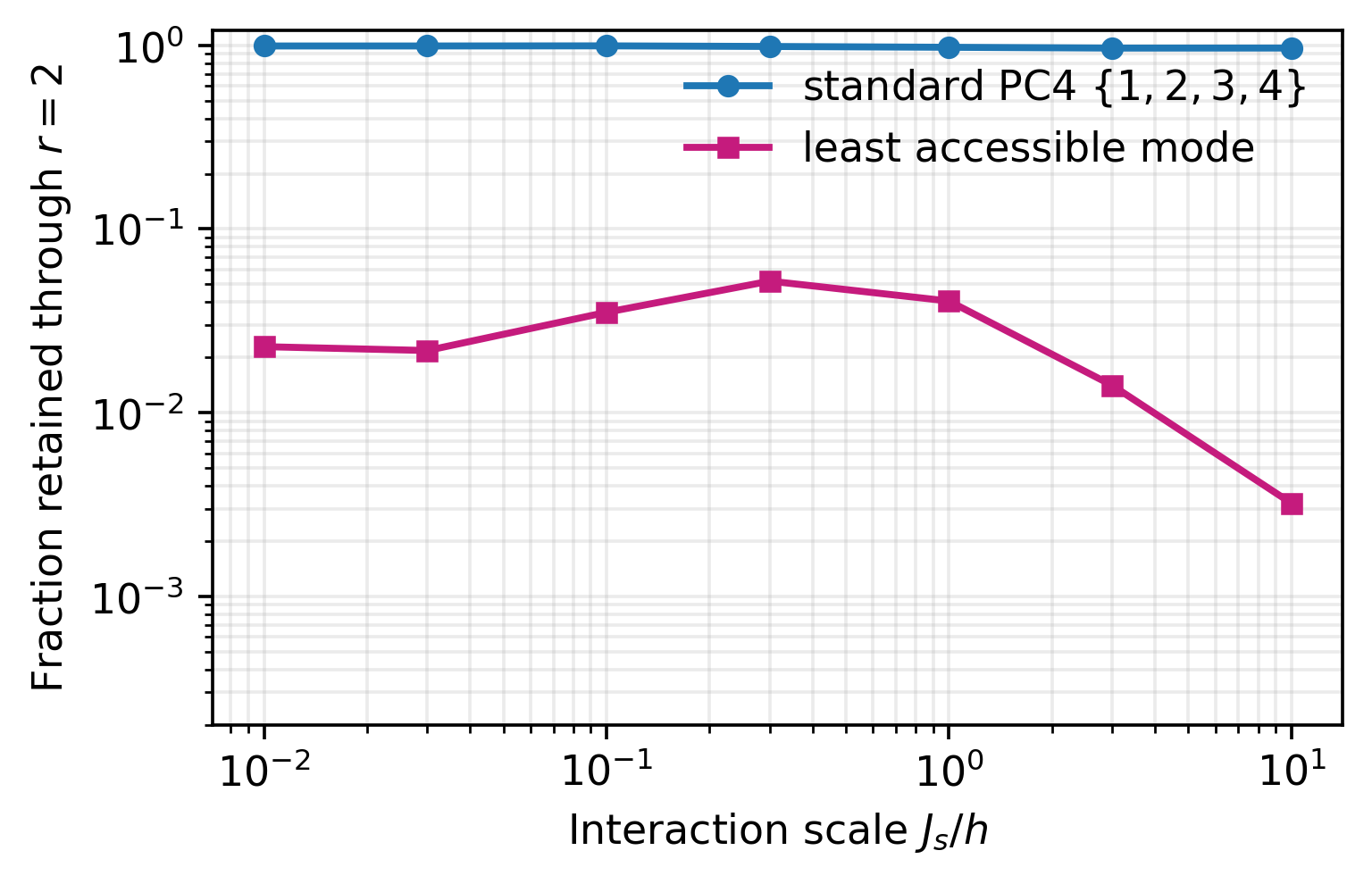}
\caption{\textbf{Task-specific accessibility versus geometric worst case.} The standard PC4 coordinate remains mostly visible in two-body moments even when another fourth-order temporal mode is almost absent from that low-order subspace. A single nonlinear benchmark can therefore underestimate the measurement complexity of the reservoir's full temporal representation.}
\label{fig:target}
\end{figure}

Figure~\ref{fig:target} compares the standard PC4 coordinate $z_tz_{t-1}z_{t-2}z_{t-3}$ with the least accessible linear combination in the same fourth-order temporal sector. The PC4 direction remains almost completely visible in moments through order two, whereas the worst direction becomes nearly invisible at strong coupling. A single parity benchmark can therefore give an overly favorable picture of the reservoir's broader temporal representation.

The worst mode is not a prediction of universal task difficulty. It identifies a temporal direction that the chosen target may not probe, while Eq.~\eqref{eq:taskcapacityscore} gives the capacity along the specified target direction.

\subsubsection{Covariance predictions agree with held-out QRC performance}
\label{sec:predexample}

We next test whether the covariance expression for task capacity predicts the
performance of an ordinarily trained QRC readout. A representative $J_s/h=1$
reservoir is driven by a stationary
i.i.d. binary stream. At each time step, the feature vector contains the
expectation values of all local-$X$ products through measured many-body order
$r$. Thus $r$ takes only the five discrete values $1,\ldots,5$.

We generate three independent input records. A reference trajectory of $10^5$
time steps is used to estimate the dynamic feature covariance
$\cC_r^{\rm dyn}$ and the feature--target covariance $\bm d_{r,y}$.
Substitution into Eq.~\eqref{eq:taskcapacityscore} gives a covariance estimate
of the optimal linear-readout capacity. Ordinary least-squares weights are
then fitted on a separate $30{,}000$-step trajectory and evaluated on a
held-out trajectory of the same length. No score bias or finite-shot
measurement noise is introduced in this comparison.

\begin{figure}[t]
\includegraphics[width=\columnwidth]{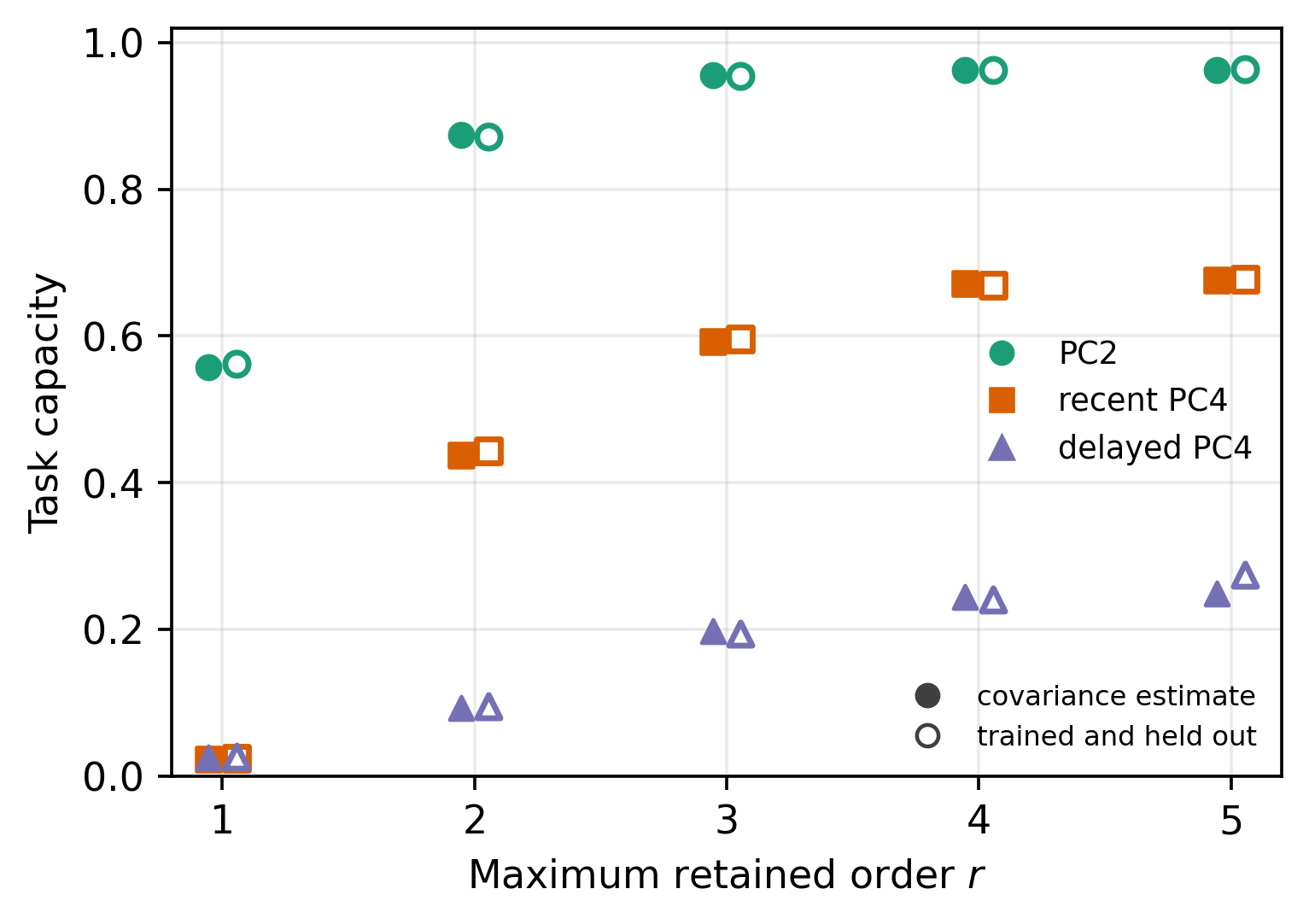}
\caption{\textbf{Covariance estimate versus trained QRC performance.}
Filled markers show the optimal linear-readout capacity estimated from an
independent labeled reference trajectory. Open markers show capacities
obtained by fitting an ordinary least-squares readout on a separate training
trajectory and evaluating it on a held-out trajectory. Small horizontal
offsets separate the paired markers visually. Results are shown only at the
five available retained correlation orders $r=1,\ldots,5$. The close
agreement of each pair shows that the covariance estimate predicts held-out
linear-readout performance.}
\label{fig:prediction}
\end{figure}

Figure~\ref{fig:prediction} compares the two results at each available order.
Filled markers show the covariance estimates, while open markers show the
independently trained and held-out capacities. Their agreement verifies that
the covariance construction reproduces ordinary linear-readout performance.
The task dependence is also clear: PC2 is already accessible from low-order
features, recent PC4 benefits substantially from including higher-body
correlations, and PC4 shifted one step further into the past remains harder.

For these benchmark targets, the task scores coincide with individual Walsh
modes, so Eq.~\eqref{eq:taskcapacityscore} gives the stationary least-squares
ceiling. The fitted readout approaches this ceiling using finite training data
but cannot recover a target component that is absent from the retained
features.

\begin{figure}[t]
\includegraphics[width=\columnwidth]{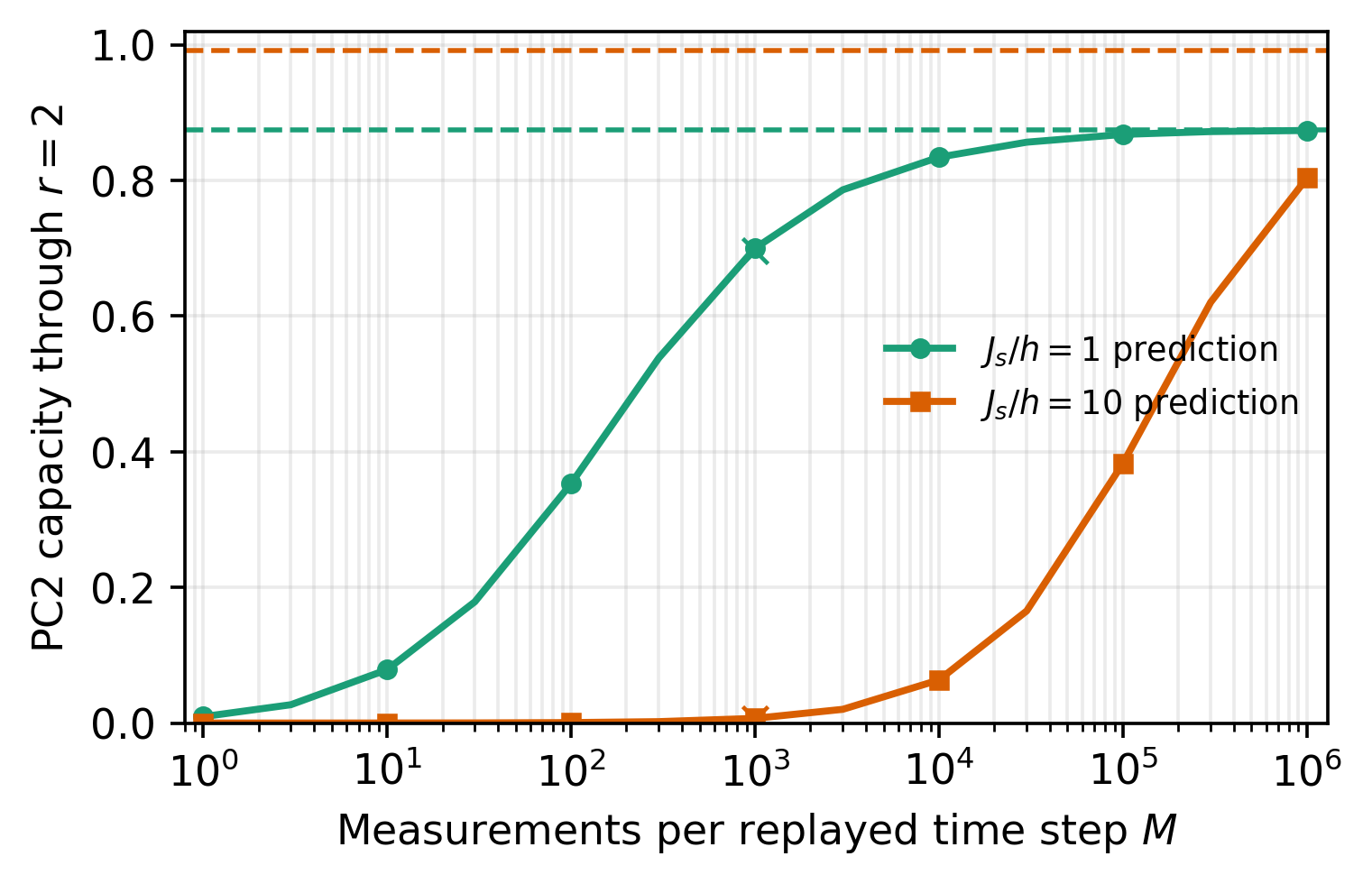}
\caption{\textbf{A measurement-budget design diagnostic.}
Stationary PC2 capacity from pair observables for representative $J_s/h=1$ and
$10$ reservoirs as a function of independent measurement shots $M$ per
replayed time step. Dashed levels are the ideal expectation-value limits;
crosses are direct finite-shot estimates from trained and held-out trajectories at $M=1000$. The
strongly interacting reservoir appears excellent in the noiseless limit but
requires far more measurements to expose the same temporal feature.}
\label{fig:shotbudget}
\end{figure}

For the finite-shot checks, independent local-$X$ bit strings are sampled from
the Born distribution of each replayed reservoir state and averaged before
fitting the readout. Figure~\ref{fig:shotbudget} compares the preceding
$J_s/h=1$ reservoir with a $J_s/h=10$ realization. Although the strongly interacting reservoir has a slightly
higher expectation-value limit, its capacity grows far more slowly with the
measurement budget. The held-out finite-shot checks follow the predicted
curves. The figure therefore demonstrates that a ranking based on exact
expectation values can reverse when the available number of measurements is
taken into account.

Figures~\ref{fig:prediction} and \ref{fig:shotbudget} test two predictions obtained from the feature covariances. In Fig.~\ref{fig:prediction}, the dynamic feature covariance $\cC_r^{\rm dyn}$ and feature--target covariance $\bm d_{r,y}$ predict the maximum linear-readout capacity for each task when correlations through order $r$ are retained. In Fig.~\ref{fig:shotbudget}, adding the finite-shot contribution $\cC_r^{\rm shot}/M$ predicts how a limited measurement budget reduces that capacity. These quantities therefore give, before readout training, both the ideal capacity at a chosen feature order and its dependence on the number of measurements. The held-out fits test these predictions for finite training records.

\subsection{Correlated signals and data-defined task coordinates}
\label{sec:correlated_numerics}

The Walsh analysis above deliberately exploits the exceptional simplicity of an independent unbiased source: its temporal basis is known analytically and orthogonal under the operating distribution. We now remove that assumption while keeping the same reservoir model and binary input levels. The source itself has temporal correlations, and the two outputs are correlated with one another, so the relevant score directions must be defined with respect to the actual stationary distribution. This second part therefore shifts from basis-resolved spectroscopy of a controlled temporal sector to task-resolved spectroscopy of a data-defined prediction problem.

\subsubsection{Correlated multi-output prediction from the operating distribution}
\label{sec:correlated_multitask}

We test the general task-score construction with a stationary correlated source and two simultaneous prediction variables. The source is a symmetric first-order Markov chain with $\Pr(z_t=z_{t-1})=p=0.8$. Its single-symbol marginal is unbiased, but successive symbols are correlated, with $m\equiv\mathbb E[z_tz_{t-1}]=2p-1=0.6$. We choose
$y_{1t}=z_tz_{t-1}$ and $y_{2t}=z_tz_{t-2}$. Their means are $m$ and $m^2$, respectively, and their nonzero covariance makes them nonorthogonal task directions.

Population Gram--Schmidt gives
$s_1=(y_1-m)/\sqrt{1-m^2}$ and
$s_2=(y_2-my_1)/\sqrt{1-m^2}$. These analytic scores provide a population reference for the calculation. The record-based calculation below instead applies the empirical construction of Appendix~\ref{app:gramschmidt}, which obtains the same two-dimensional target subspace directly from labeled data.

We use one fixed $J_s/h=1$ reservoir. We first calculate the state-level quantities directly from the known Markov transition probabilities and reservoir channels. Let $a=z_t$, $b=z_{t-1}$, and $c=z_{t-2}$. Define the unnormalized stationary conditional operator
\begin{equation}
R_a
\equiv
P_{\rm ref}(z_t=a)\,
\mathbb E[\rho_t\mid z_t=a],
\qquad
\Tr R_a=P_{\rm ref}(a).
\label{eq:markov_conditional_operator}
\end{equation}
Using the reservoir update in Eq.~\eqref{eq:qrcupdate}, the first-order Markov property, and stationarity gives the coupled fixed-point equations
\begin{equation}
R_a=\sum_b P(a|b)\Phi_a(R_b).
\label{eq:markov_conditional_fixedpoint}
\end{equation}
Solving these equations avoids estimating the state-level geometry from a finite trajectory.

Both targets are determined by the three-symbol history $(a,b,c)$.
For each three-symbol history $(a,b,c)$, define the normalized conditional reservoir state
\begin{equation}
\bar\rho_{abc}
\equiv
\mathbb E\!\left[
\rho_t
\,\middle|\,
z_t=a,\,
z_{t-1}=b,\,
z_{t-2}=c
\right].
\end{equation}
The corresponding unnormalized history-resolved state is
\begin{align}
&T_{abc}
\equiv
P_{\rm ref}(z_t=a,z_{t-1}=b,z_{t-2}=c)\,
\bar\rho_{abc}
\nonumber\\
&=
P(z_t=a\mid z_{t-1}=b)\,
\Phi_a\!\left[
P(z_{t-1}=b\mid z_{t-2}=c)\,
\Phi_b(R_c)
\right].
\label{eq:three_symbol_operator}
\end{align}
Here,
\begin{equation}
R_c
=
P_{\rm ref}(z_{t-2}=c)\,
\mathbb E[\rho_{t-2}\mid z_{t-2}=c]
\end{equation}
is the unnormalized stationary reservoir state associated with the input value $c$.

Because the channels are trace preserving, we have
\begin{equation}
\Tr T_{abc}
=P_{\rm ref}(z_t=a,z_{t-1}=b,z_{t-2}=c),
\label{eq:three_symbol_joint_state}
\end{equation}
which will be denoted for simplicity as
$P_{\rm ref}(a,b,c)$ in the following.

Substituting $\bm z=(a,b,c)$ into the score ensemble of Eq.~\eqref{eq:scoreensemble} gives
\begin{align}
\rho(\bm\theta)
&=
\sum_{a,b,c}
P_{\bm\theta}(a,b,c)\bar\rho_{abc}
\nonumber\\
&=
\sum_{a,b,c}T_{abc}
+
\sum_A\theta_A
\sum_{a,b,c}s_A(a,b,c)T_{abc}.
\label{eq:three_symbol_affine_expansion}
\end{align}
Comparison with the general affine state in Eq.~\eqref{eq:affine} therefore identifies
\begin{equation}
\rho_{\rm ref}
=
\sum_{a,b,c}T_{abc},
\qquad
\Gamma_A
=
\sum_{a,b,c}s_A(a,b,c)T_{abc}.
\label{eq:three_symbol_reference_tangents}
\end{equation}
Thus the unweighted sum gives the stationary state, whereas the score-weighted sums give the two task tangents.

\begin{figure}[t]
\centering
\includegraphics[width=\columnwidth]{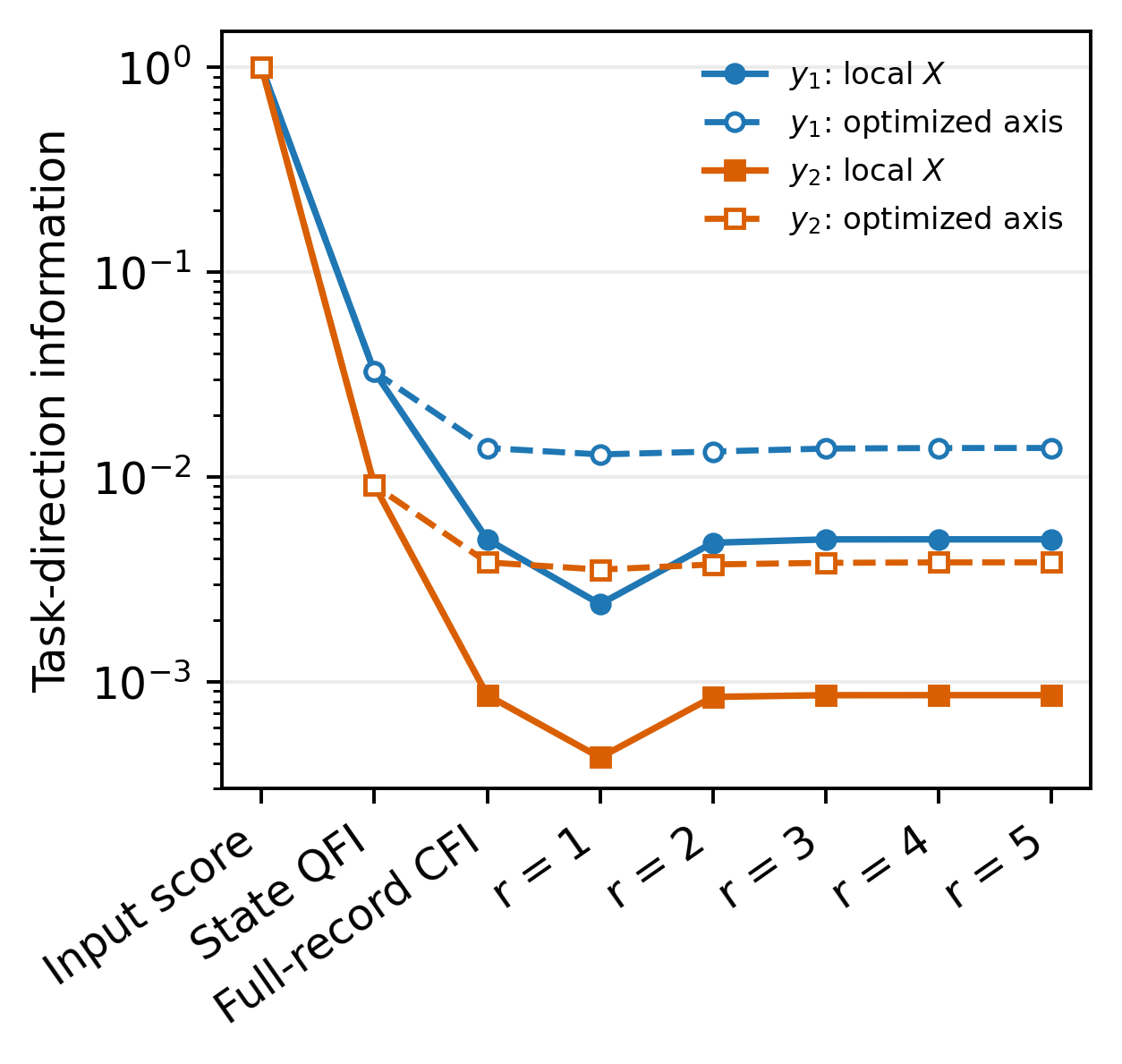}
\caption{\textbf{Task-resolved information hierarchy for correlated multi-output prediction.} A stationary binary Markov source drives one fixed $J_s/h=1$ reservoir. Each task score has unit variance, so its curve starts at one and then shows the reservoir QFI, the complete-record CFI for one measurement shot, and the information retained by products through many-body order $r$. Solid lines use the local-$X$ measurement; dashed lines use the optimized common measurement axis introduced below.}
\label{fig:corrhier}
\end{figure}

From $\rho_{\rm ref}$ and $\Gamma_A$ we calculate the state QFIM $\cH_{\rm task}$. For a complete local-$X$ measurement, the same operators give the outcome probabilities and their score derivatives, and hence the CFIM $\cF$ from Eq.~\eqref{eq:cfi}. Compressing each local-$X$ bit string to products through order $r$ gives $\cB_r$ from Eq.~\eqref{eq:Br}. For each original output, we evaluate these matrices along its normalized direction in the score basis, as in Eq.~\eqref{eq:taskcapacitymulti}. Figure~\ref{fig:corrhier} displays the resulting quadratic forms from the normalized input score to the state QFI, complete-record CFI, and truncated moment information.

For both outputs, Fig.~\ref{fig:corrhier} shows that most of the loss occurs before or at measurement: first between the input score and the reservoir state, and then between the state and the local-$X$ record. Once that bit string is available, moments through order two retain nearly all of its task information. The first output remains more accessible than the second throughout the hierarchy. The dashed curves show that the optimized measurement axis considered below recovers part of the state-to-measurement loss.

We next calculate the capacities relevant to an ordinary QRC readout using exact expectation-value features. A labeled reference trajectory of $10^5$ time steps supplies the empirical Gram--Schmidt scores, the feature--score response matrix $\cD_r$, and the dynamic feature covariance $\cC_r^{\rm dyn}$ for each $r=1,\ldots,5$. Substituting these estimates into $\cP_r^{(\infty)}$ in Eq.~\eqref{eq:Plimits} and then using Eq.~\eqref{eq:finitebudgetcapacity} gives the predicted capacity of each original output. We then fit ordinary least-squares readouts on an independent $30{,}000$-step trajectory and evaluate them on a separate trajectory of the same length.

\begin{figure}[t]
\centering
\includegraphics[width=\columnwidth]{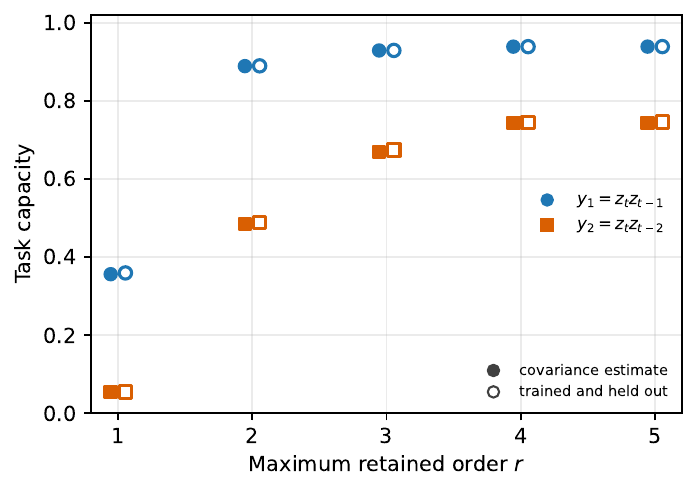}
\caption{\textbf{Covariance prediction for correlated multi-output QRC.} Filled markers show capacities calculated from a labeled stationary calibration record, and open markers show capacities from independently trained and held-out linear readouts. Circles and squares denote the two original, nonorthogonal outputs. Small horizontal offsets separate the paired markers at each available retained order $r=1,\ldots,5$. The first output approaches its plateau at lower order, whereas the second continues to benefit from higher-body features. The close agreement within each pair shows that the covariance calculation predicts ordinary linear-readout performance.}
\label{fig:corrvalidation}
\end{figure}

Figure~\ref{fig:corrvalidation} compares the two capacities directly at each retained order $r=1,\ldots,5$. Filled markers show the covariance estimates from the labeled calibration record, while open markers show the independently trained and held-out capacities. The paired markers agree closely for both outputs at every order. The first output gains most of its capacity when two-body features are included and is already close to its plateau by $r=3$. The second output remains less accessible at every order and continues to benefit from higher-body measured correlations through $r=4$. Thus two targets of the same nonlinear order can have different readout requirements when they probe different delays. The agreement of the paired markers shows that the empirical score construction preserves these original, correlated targets and that the resulting covariances predict their ordinary linear-readout performance.

\begin{figure}[t]
\centering
\includegraphics[width=\columnwidth]{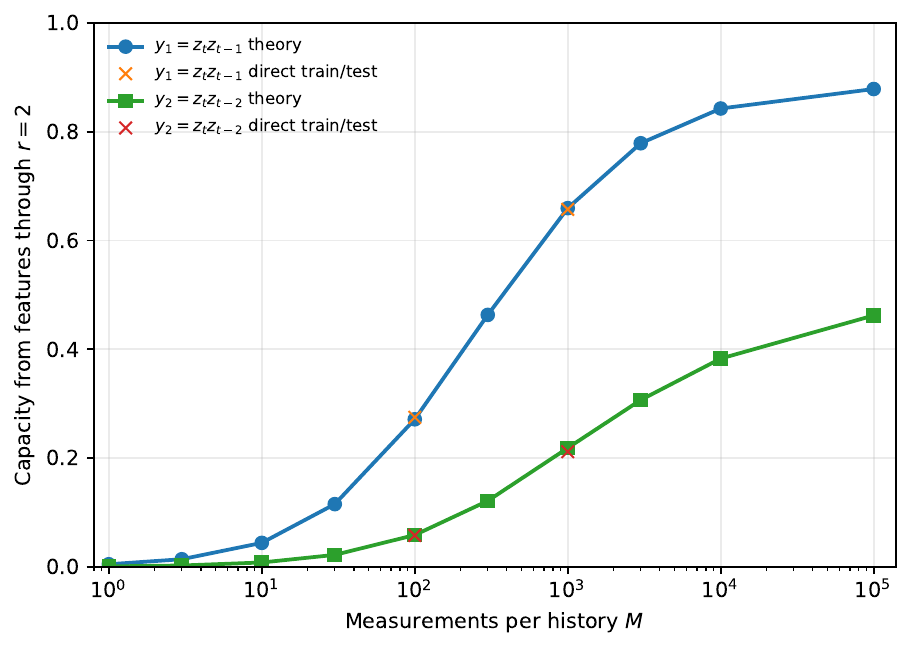}
\caption{\textbf{Finite measurement budget for correlated outputs.} Task capacities from local-$X$ products through $r=2$ versus the number $M$ of independent measurements of each history-conditioned state. Curves evaluate Eq.~\eqref{eq:finitebudgetcapacity}; crosses are direct finite-shot train/test checks at $M=100$ and $1000$.}
\label{fig:corrbudget}
\end{figure}

Finally, we calculate how a finite number of measurements changes the prediction capacity. For features through $r=2$, the reference calculation separates the dynamic covariance from the one-shot measurement covariance and forms $\cC_2(M)$ using Eq.~\eqref{eq:CM}. The response matrix is independent of $M$ by Eq.~\eqref{eq:DindependentM}; inserting $\cC_2(M)$ into Eqs.~\eqref{eq:PM} and \eqref{eq:finitebudgetcapacity} gives the two capacity curves in Fig.~\ref{fig:corrbudget}. For direct checks, finite-shot feature vectors are sampled from the local-$X$ Born probabilities on independent training and test trajectories, after which the readout is fitted and evaluated in the usual way.

Figure~\ref{fig:corrbudget} shows that both capacities increase with the measurement budget and approach their exact-expectation limits. The second output rises more slowly and remains the harder prediction problem. The direct train/test points follow the covariance curves, showing that the same task coordinates also predict the measurement resources required by the two outputs.

\subsubsection{Task-guided optimization of the measurement protocol}
\label{sec:measurement_optimization}

The task-resolved hierarchy can be used not only to diagnose where predictive information is lost, but also to optimize the measurement itself. We demonstrate this for the correlated two-target example by keeping the reservoir Hamiltonian, dissipation, input statistics, and task-score subspace fixed, and varying only a common local spin-measurement axis
\begin{equation}
\bm n(\vartheta,\varphi)
=
(\sin\vartheta\cos\varphi,
 \sin\vartheta\sin\varphi,
 \cos\vartheta).
\end{equation}
Each spin is measured in the eigenbasis of
$\sigma_{\bm n}=n_xX+n_yY+n_zZ$. For a complete outcome string
$\bm x=(x_1,\ldots,x_N)$, the corresponding POVM element is
$M_{\bm x}(\bm n)=\bigotimes_{i=1}^{N}\frac{I+x_i\sigma_{\bm n}}{2}$. At fixed reservoir dynamics, the reference state $\rho_{\rm ref}$ and task tangents $\Gamma_A$ do not change when the measurement axis is varied. Hence each trial axis gives
$p_{\bm x}(\bm n)=\Tr[M_{\bm x}(\bm n)\rho_{\rm ref}]$ and
$\dot p_{\bm xA}(\bm n)=\Tr[M_{\bm x}(\bm n)\Gamma_A]$, from which the complete-record Fisher matrix $\cF_{\bm n}$ (subscript $\bm n$ indicates the measurement direction) and the truncated moment matrices $\cB_r(\bm n)$ are evaluated exactly as in Sec.~\ref{sec:spectroscopy}.

Because
$0\preceq\cB_r(\bm n)\preceq\cF_{\bm n}\preceq\cH_{\rm task}$,
we normalize each measurement-level matrix by the state-level QFIM before comparing different measurement protocols. Specifically,
$\cH_{\rm task}^{-1/2}\cB_r(\bm n)\cH_{\rm task}^{-1/2}$
expresses the retained information relative to the state-level task metric. For the present two-output problem, we define the normalized order-$r$ task accessibility as the mean of its two eigenvalues,
\begin{equation}
\eta_r(\bm n)
=
\frac{1}{2}\Tr\!\left[
\cH_{\rm task}^{-1/2}
\cB_r(\bm n)
\cH_{\rm task}^{-1/2}
\right].
\label{eq:eta_measurement}
\end{equation}
The two eigenvalues give the fractions of the available state-level task information retained along the corresponding task directions, while $\eta_r$ gives their average. 

This suggests an order-aware measurement design: if an experiment can reliably retain measured correlations only through order $r_{\max}$, the appropriate protocol is
\begin{equation}
\bm n_{\rm opt}^{(r_{\max})}
=
\arg\max_{\bm n}\eta_{r_{\max}}(\bm n),
\label{eq:nopt_measurement}
\end{equation}
rather than an axis optimized for the complete record but not for the experimentally usable feature set.

\begin{figure}[t]
\includegraphics[width=\columnwidth]{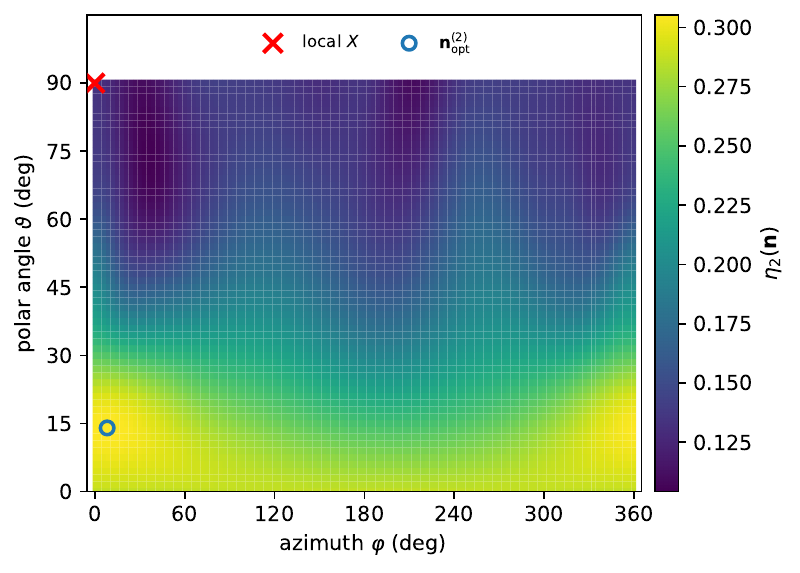}
\caption{\textbf{Task-guided optimization of a common local measurement axis.}
Normalized order-two task accessibility $\eta_2(\bm n)$ over the upper Bloch hemisphere for the fixed correlated-source reservoir realization. The antipodal axes $\bm n$ and $-\bm n$ differ only by relabeling the binary outcomes. Markers identify local $X$ and the optimized order-two axis near $+Z$.}
\label{fig:measurement_landscape}
\end{figure}

For the fixed five-spin realization, we numerically maximize $\eta_r(\bm n)$ over the two Bloch-sphere angles and confirm the resulting optimum on an independent angular grid. Figure~\ref{fig:measurement_landscape} shows the order-two objective over one hemisphere. The optimum lies close to the $+Z$ axis, while local $X$ is far from the high-accessibility region. For this task and reservoir, adapting the measurement axis therefore recovers substantially more state-level task information than local $X$.

\begin{figure}[t]
\includegraphics[width=\columnwidth]{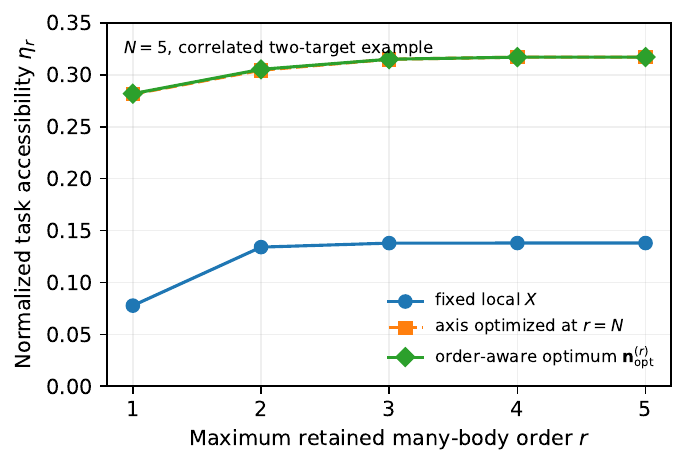}
\caption{\textbf{Measurement optimization adapted to the experimentally retained many-body order.}
The normalized task accessibility $\eta_r$ defined in Eq.~\eqref{eq:eta_measurement} is shown versus the maximum retained correlation order $r$.
The fixed local-$X$ measurement is compared with a single axis optimized for the complete record ($r=N$) and with the order-aware optimum $\bm n_{\rm opt}^{(r)}$ obtained by maximizing $\eta_r(\bm n)$ over the measurement axis separately at each $r$. In this realization, the complete-record optimum is already close to the order-specific optima, but both strongly outperform local $X$.}
\label{fig:measurement_optimization_curve}
\end{figure}

Figure~\ref{fig:measurement_optimization_curve} compares local $X$, an axis optimized for the complete record, and an order-aware optimum. Optimized measurements outperform local $X$ at every retained order, with the largest relative gain when only low-order features are available. In this realization, the complete-record optimum nearly coincides with the order-specific optima, so one optimized axis performs well across the entire feature hierarchy.

\section{Conclusion}
\label{sec:conclusion}

We have introduced task-resolved Fisher spectroscopy as a task-specific characterization of quantum reservoir computing at its stationary operating distribution. Instead of defining information coordinates through an abstract perturbation of a physical control parameter, the prediction targets themselves define normalized score directions on the input histories. For several correlated outputs, orthonormalizing the centered targets gives orthogonal coordinates spanning the same prediction subspace. Reweighting labeled histories along these scores then produces an exactly affine family of reservoir states and measurement statistics. The state-level quantum Fisher matrix, the classical Fisher matrix of a chosen measurement, and the information retained by measured correlations of increasing many-body order can therefore be compared in the same task-defined coordinates.

The central result is the information hierarchy in Eq.~\eqref{eq:mainhier}. It distinguishes task information hidden by the chosen measurement from information discarded when the measured record is compressed to a restricted feature set. The normalized task scores also provide a direct reference for assessing reservoir encoding. At the same time, the quadratic form of the truncated moment matrix is exactly the optimal stationary linear-readout capacity for the corresponding target. This identity connects the Fisher hierarchy directly to the performance measure used in QRC, rather than to an unrelated diagnostic. The finite-measurement-budget extension further separates variations of the reservoir response across input histories from measurement shot noise and predicts how the usable capacity approaches the exact-expectation limit as the number of measurements increases. These measurement-level quantities can be obtained from stationary labeled records and measured outcome strings without an analytic model of the input distribution or quantum-state tomography; only the state-level quantum Fisher matrix generally requires additional state characterization.

The open-spin calculations demonstrate the complementary roles of the framework. For independent unbiased inputs, the Walsh construction resolves an entire temporal subspace and shows that interactions can move task information into higher-body measured correlations, making low-order readout statistically costly. It also shows why the capacity of a single parity target need not reveal the least accessible temporal direction, and why reservoirs ranked by exact expectation values can be ranked differently at finite measurement budget. For a correlated Markov source, task coordinates constructed directly from labeled data predict the held-out capacities of two correlated outputs, including their different requirements for retained correlation order and measurement shots. Finally, optimizing the local measurement axis within the same task subspace recovers information that is poorly exposed by a fixed local-$X$ measurement. Task-resolved Fisher spectroscopy therefore serves not only as a diagnostic of a trained architecture, but also as a quantitative objective for choosing how the reservoir is measured and how its output is represented.

The construction is not tied to the five-spin model, binary independent inputs, or a particular local measurement. A natural next step is to extend the data-defined score construction to continuous and multivariate signals, larger output spaces, and more general measurement protocols. For larger reservoirs, scalable estimates of the required response and covariance matrices will be important when complete outcome histograms or all many-body correlators cannot be retained. The same hierarchy can also be used to optimize reservoir controls, measurement settings, retained feature order, and shot allocation jointly under an experimental resource constraint. Such developments would turn QRC assessment from a comparison of isolated benchmark scores into a task- and resource-resolved design procedure for physical information processors.

\begin{acknowledgments}
This work is supported by the U.S. National Science Foundation under Grants No.\ PHY-2216774 and No.\ DMR-2406524.
\end{acknowledgments}
\appendix

\section{Exact Walsh-history identity and response}\label{app:walsh}
Let $z_1$ denote the most recent symbol of a length-$L$ input block, so that the block propagator is
\begin{equation}
\mathcal T_{\bm z}=\Phi_{z_1}\circ\Phi_{z_2}\circ\cdots\circ\Phi_{z_L}.
\end{equation}
Using $\Phi_z=\PhiB+z\Delta$, expand the product of channels by choosing either $\PhiB$ or $z_k\Delta$ at every delay. This gives
\begin{equation}
\mathcal T_{\bm z}(\rhoB)=\sum_{B\subseteq\{1,\ldots,L\}}\chi_B(\bm z)\,\Gamma_B,
\label{eq:appendix_channel_expansion}
\end{equation}
where $\chi_\emptyset=1$, $\Gamma_\emptyset=\PhiB^L(\rhoB)=\rhoB$, and, for $B=\{d_1<\cdots<d_q\}$,
\begin{align}
\Gamma_B={}&\PhiB^{d_1-1}\Delta\,
\PhiB^{d_2-d_1-1}\Delta\cdots\nonumber\\
&\times\PhiB^{d_q-d_{q-1}-1}\Delta(\rhoB).
\label{eq:appendix_gamma}
\end{align}
The powers of $\PhiB$ associated with delays older than $d_q$ drop out because $\PhiB(\rhoB)=\rhoB$.

The biased history distribution is
\begin{equation}
P_{\bm\theta}(\bm z)=2^{-L}\left[1+\sum_{A}\theta_A\chi_A(\bm z)\right].
\end{equation}
Averaging Eq.~\eqref{eq:appendix_channel_expansion} over histories gives
\begin{align}
\rho(\bm\theta)
&=\sum_{\bm z}P_{\bm\theta}(\bm z)\mathcal T_{\bm z}(\rhoB)\nonumber\\
&=\sum_B\Gamma_B\,2^{-L}\sum_{\bm z}\chi_B(\bm z)\nonumber\\
&\quad+\sum_{A,B}\theta_A\Gamma_B\,2^{-L}
\sum_{\bm z}\chi_A(\bm z)\chi_B(\bm z).
\end{align}
The Walsh modes are orthonormal under the uniform reference distribution,
\begin{equation}
2^{-L}\sum_{\bm z}\chi_A(\bm z)\chi_B(\bm z)=\delta_{AB},
\end{equation}
and every nonconstant mode has zero uniform mean. Hence only $B=\emptyset$ survives from the unbiased term and exactly one channel subset survives for each $\theta_A$:
\begin{equation}
\rho(\bm\theta)=\rhoB+\sum_A\theta_A\Gamma_A.
\end{equation}
There are no higher powers of $\bm\theta$. This is the explicit channel realization of the general affine score ensemble for the stationary i.i.d. unbiased binary construction used in the numerical study.

\section{Projector proof of the moment hierarchy}\label{app:projector}
Here we prove the moment part of Eq.~\eqref{eq:mainhier} and explain the meaning of the matrix inequalities. All outcome-space vectors and sums below are understood on the support $p_x>0$, as in Eq.~\eqref{eq:cfi}.

Introduce the response matrix
\begin{equation}
S_{xA}=\frac{\dot p_{xA}}{\sqrt{p_x}},
\end{equation}
where $A$ labels a task-score direction. The Fisher matrix of the complete measurement record is then
\begin{equation}
\cF=S^TS.
\end{equation}
For the features retained through order $r$, define the centered functions
$g_{r,\alpha}(x)=f_{r,\alpha}(x)-\langle f_{r,\alpha}\rangle$
and the outcome-space matrix
\begin{equation}
(\mathsf G_r)_{x\alpha}=\sqrt{p_x}\,g_{r,\alpha}(x).
\end{equation}
The feature covariance and response matrices can then be written as
\begin{equation}
\cC_r=\mathsf G_r^T\mathsf G_r,
\qquad
\cD_r=\mathsf G_r^TS.
\end{equation}

Now define
\begin{equation}
\Pi_r
=
\mathsf G_r
(\mathsf G_r^T\mathsf G_r)^+
\mathsf G_r^T.
\end{equation}
The Moore--Penrose pseudoinverse ensures that this expression remains valid even when some retained features are linearly dependent. The matrix $\Pi_r$ is symmetric and satisfies $\Pi_r^2=\Pi_r$; it is therefore the orthogonal projector onto the space spanned by the columns of $\mathsf G_r$. In other words, $\Pi_r$ keeps the part of an outcome-dependent response that can be represented by the retained features and removes the orthogonal part. Substitution into Eq.~\eqref{eq:Br} gives
\begin{equation}
\cB_r
=
\cD_r^T\cC_r^+\cD_r
=
S^T\Pi_rS.
\label{eq:projector}
\end{equation}

This projector form makes the ordering with $r$ explicit. Every feature retained at order $r$ is also retained at order $r+1$. Writing $\mathrm{col}(\mathsf G_r)$ for the space spanned by the columns of $\mathsf G_r$, we therefore have
\begin{equation}
\mathrm{col}(\mathsf G_r)
\subseteq
\mathrm{col}(\mathsf G_{r+1}).
\end{equation}
For two nested subspaces, $\Pi_{r+1}-\Pi_r$ is itself the orthogonal projector onto the newly added directions. Let $\bm v$ be any real vector specifying a linear combination of the task-score directions. The outcome-space vector $S\bm v$ is the normalized change of the complete outcome probabilities along that combined task perturbation. Equation~\eqref{eq:projector} gives
\begin{align}
\bm v^T(\cB_{r+1}-\cB_r)\bm v
&=
(S\bm v)^T(\Pi_{r+1}-\Pi_r)(S\bm v)
\nonumber\\
&=
\left\|(\Pi_{r+1}-\Pi_r)S\bm v\right\|^2
\ge 0.
\label{eq:projector_increment}
\end{align}
Thus adding higher-order features cannot reduce the accessible information along any task direction.

A real symmetric matrix $M$ is called positive semidefinite when
$\bm v^TM\bm v\ge0$ for every $\bm v$. The notation
$\cB_r\preceq\cB_{r+1}$, often called the Löwner ordering, means precisely that $\cB_{r+1}-\cB_r$ is positive semidefinite. It is not an element-by-element comparison of the two matrices. Equation~\eqref{eq:projector_increment} therefore proves
\begin{equation}
0\preceq\cB_1\preceq\cB_2\preceq\cdots\preceq\cB_N.
\end{equation}
The initial inequality follows in the same way from
$\bm v^T\cB_r\bm v=\|\Pi_rS\bm v\|^2\ge0$.

It remains to identify the last member of this sequence. For a measurement of $N$ binary variables, the $2^N$ product functions
$f_Q(x)=\prod_{i\in Q}x_i$, including the constant function for $Q=\emptyset$, form a complete basis for all functions of the measured bit string. After centering, the $2^N-1$ nonconstant products therefore span the full zero-mean outcome space. The measurement probabilities remain normalized under every task perturbation, so
\begin{equation}
\sum_x\dot p_{xA}=0
\quad\Longrightarrow\quad
\sum_x\sqrt{p_x}\,S_{xA}=0.
\end{equation}
Each column of $S$ consequently lies in the zero-mean subspace spanned by the complete feature matrix $\mathsf G_N$. Hence $\Pi_NS=S$, and
\begin{equation}
\cB_N=S^T\Pi_NS=S^TS=\cF.
\end{equation}
Combining the results gives the moment-information hierarchy
\begin{equation}
0\preceq\cB_1\preceq\cdots\preceq\cB_N=\cF.
\end{equation}

The remaining inequality in Eq.~\eqref{eq:mainhier} has the same meaning: it holds for every task combination $\bm v$. A chosen quantum measurement cannot contain more Fisher information than the reservoir state, giving $\cF\preceq\cH_{\rm task}$. Equivalently, the full hierarchy states that, for every $\bm v$,
\begin{equation}
0
\le
\bm v^T\cB_r\bm v
\le
\bm v^T\cF\bm v
\le
\bm v^T\cH_{\rm task}\bm v.
\end{equation}

\section{Response--covariance identity for a general score ensemble}
\label{app:response_covariance}

Here we derive Eq.~\eqref{eq:Drdef} for the general operating distribution. Let $q(x|\bm z)$ denote the probability of measurement outcome $x$ conditioned on the recent history window $\bm z$, with any earlier reservoir context averaged according to the stationary reference process. Under Eq.~\eqref{eq:scoreensemble}, the joint distribution is $P_{\bm\theta}(\bm z,x)=P_{\bm\theta}(\bm z)q(x|\bm z)$. For a measured feature $f_\alpha(x)$,
\begin{align}
\left.\frac{\partial}{\partial\theta_A}\mathbb E_{\bm\theta}[f_\alpha]\right|_{\bm0}
&=\sum_{\bm z,x}P_{\rm ref}(\bm z)s_A(\bm z)q(x|\bm z)f_\alpha(x)\nonumber\\
&=\mathbb E_{\rm ref}[f_\alpha s_A].
\end{align}
Since $\mathbb E_{\rm ref}[s_A]=0$, the last expression is $\Cov_{\rm ref}(f_\alpha,s_A)$, proving Eq.~\eqref{eq:Drdef}.

Equivalently, the statistical score of the family is $\left.\partial_{\theta_A}\ln P_{\bm\theta}(\bm z)\right|_{\bm0}=s_A(\bm z)$, so Eq.~\eqref{eq:Drdef} is the standard score-function response identity. The construction is exact because the reweighted distribution is linear in $\bm\theta$.

For the task score $s_y=(y-\mu_y)/\sigma_y$, this immediately gives $\bm D_{r,y}=\Cov_{\rm ref}(\bm f_r,y)/\sigma_y$. Substitution into Eq.~\eqref{eq:explainedvariance} yields Eq.~\eqref{eq:taskcapacityscore}. Thus the equality between task-directed moment information and linear-readout capacity does not rely on an unbiased input distribution or on a Walsh expansion.

The same identity holds for empirical reweighting. For sampled segments $\bm z_t^{(n)}$ with centered empirical scores $\widehat s_A^{(n)}$, the weights $w_n(\theta_A)=1+\theta_A\widehat s_A^{(n)}$ satisfy $\sum_n w_n=N_s$. Hence
\begin{equation}
\frac{\widehat{\bm\mu}_r(+\epsilon)-\widehat{\bm\mu}_r(-\epsilon)}{2\epsilon}
=\frac1{N_s}\sum_{n=1}^{N_s}\widehat s_A^{(n)}\bm f_r(\bm x_t^{(n)}),
\end{equation}
which is the empirical response--covariance estimator used in Sec.~\ref{sec:experiment}. Consecutive windows from one long trajectory need not be independent for consistency of stationary averages, although their correlations must be retained when estimating statistical uncertainties.

\section{Derivation of the generalized-mode sampling overhead}
\label{app:finiteshot}
The generalized eigenvalues in Eq.~\eqref{eq:geneig} can be given a finite-sample interpretation by treating a deliberately imposed score bias as an unknown parameter after it has passed through the reservoir. This is an interpretation of the Fisher information, not an additional step required to run the spectroscopy. In the derivation below, the reference mean $\bm\mu_0$, covariance $\cC$, and response matrix $\cD$ are taken as known from separate reference measurements; uncertainty from estimating those quantities is not included in the sampling variance derived here.

For $R$ independent spectroscopy runs, let
\begin{equation}
\overline{\bm f}=\frac1R\sum_{s=1}^{R}\bm f^{(s)}.
\end{equation}
The affine response law gives
\begin{equation}
\mathbb E_{\bm\theta}[\overline{\bm f}]
=\bm\mu_0+\cD\bm\theta,
\end{equation}
and, at the reference ensemble,
\begin{equation}
\Cov_{\rm ref}(\overline{\bm f})=\frac1R\cC.
\end{equation}
Assume first that $\cC$ and $\cB=\cD^T\cC^{-1}\cD$ are nonsingular. We estimate the imposed score amplitudes by the covariance-weighted least-squares fit
\begin{equation}
\widehat{\bm\theta}
=\arg\min_{\bm\theta'}
(\overline{\bm f}-\bm\mu_0-\cD\bm\theta')^T
\cC^{-1}
(\overline{\bm f}-\bm\mu_0-\cD\bm\theta').
\end{equation}
The normal equations give
\begin{equation}
\widehat{\bm\theta}
=\cB^{-1}\cD^T\cC^{-1}
(\overline{\bm f}-\bm\mu_0).
\label{eq:app_thetaest}
\end{equation}
The hat denotes the value inferred from the finite set of measured outputs. Using the affine mean shows $\mathbb E_{\bm\theta}[\widehat{\bm\theta}]=\bm\theta$, and at $\bm\theta=0$,
\begin{align}
\Cov_{\rm ref}(\widehat{\bm\theta})
&=\cB^{-1}\cD^T\cC^{-1}
\frac{\cC}{R}
\cC^{-1}\cD\cB^{-1}\\
&=\frac1R\cB^{-1}.
\label{eq:app_thetacov}
\end{align}
No Gaussian approximation or large-$R$ limit is used in this covariance identity.

For comparison, define the score vector of the complete measured bit string by
\begin{equation}
\ell_A(x)=\frac{\dot p_{xA}}{p_x}.
\end{equation}
At the reference ensemble, $\mathbb E_{\rm ref}[\bm\ell]=0$ and $\Cov_{\rm ref}(\bm\ell)=\cF$. Affinity of $p_x(\bm\theta)$ also gives $\mathbb E_{\bm\theta}[\bm\ell]=\cF\bm\theta$. Therefore the estimator
\begin{equation}
\widehat{\bm\theta}_X
=\cF^{-1}\left(\frac1R\sum_{s=1}^{R}\bm\ell(x_s)\right)
\end{equation}
has
\begin{equation}
\Cov_{\rm ref}(\widehat{\bm\theta}_X)=\frac1R\cF^{-1}.
\end{equation}

Let the columns of $V$ be the generalized eigenvectors of Eq.~\eqref{eq:geneig}, chosen so that
\begin{equation}
V^T\cF V=I,
\qquad
V^T\cB V=\Lambda
=\operatorname{diag}(\lambda_1,\ldots,\lambda_K).
\end{equation}
Writing the temporal perturbation as $\bm\theta=V\bm\phi$ defines scalar amplitudes $\phi_j$ along the generalized temporal modes. Let $\widehat\phi_{j,X}$ and $\widehat\phi_{j,r}$ denote the corresponding estimates obtained from the complete bit string and from correlation features through order $r$, respectively. Transforming Eqs.~\eqref{eq:app_thetacov} and the full-record covariance to the $\bm\phi$ coordinates gives
\begin{equation}
\Var_{\rm ref}(\widehat\phi_{j,X})=\frac1R,
\qquad
\Var_{\rm ref}(\widehat\phi_{j,r})=\frac1{R\lambda_j^{(r)}}.
\end{equation}
Hence the variance ratio, and therefore the number-of-runs overhead for equal variance, is exactly $1/\lambda_j^{(r)}$ at the reference point.

If some measured features are exactly linearly dependent, $\cC^{-1}$ is replaced by the Moore--Penrose pseudoinverse $\cC^+$. A temporal combination with zero eigenvalue of $\cB$ produces no change in the retained feature means and therefore cannot be estimated from those features. A direction with zero eigenvalue of $\cF$ is absent even from the complete measured bit string and is excluded from the generalized-eigenvalue comparison.

\section{Gram--Schmidt construction of multiple task scores}
\label{app:gramschmidt}

Consider $K$ prediction variables $y_A=g_A(\bm z)$ defined on the same stationary input-history distribution. Let $\widetilde y_A=y_A-\mathbb E_{\rm ref}[y_A]$ and define the inner product $\langle u,v\rangle_{\rm ref}=\mathbb E_{\rm ref}[uv]$. Assuming the target functions contain $K'$ linearly independent directions, Gram--Schmidt constructs an orthonormal basis of their span. Set $u_1=\widetilde y_1$ and $s_1=u_1/\sqrt{\langle u_1,u_1\rangle_{\rm ref}}$. Recursively, for $A>1$,
\begin{equation}
u_A=\widetilde y_A-\sum_{B<A}\langle \widetilde y_A,s_B\rangle_{\rm ref}s_B,
\qquad
s_A=\frac{u_A}{\sqrt{\langle u_A,u_A\rangle_{\rm ref}}},
\label{eq:app_gramschmidt}
\end{equation}
whenever $\langle u_A,u_A\rangle_{\rm ref}>0$. A vanishing residual norm means that $y_A$ adds no new direction beyond the preceding targets and it is omitted from the score family. The retained scores satisfy $\mathbb E_{\rm ref}[s_A]=0$ and $\mathbb E_{\rm ref}[s_As_B]=\delta_{AB}$.

Every centered original target remains in the same span and has coefficients $c_B^{(A)}=\langle \widetilde y_A,s_B\rangle_{\rm ref}$, so $\widetilde y_A=\sum_B c_B^{(A)}s_B$ and $\Var_{\rm ref}(y_A)=\sum_B[c_B^{(A)}]^2$. These coefficients are the quantities used in Eq.~\eqref{eq:taskcapacitymulti} to recover the capacity of the original, generally correlated, prediction variables from the orthonormal Fisher coordinates. The order chosen for Gram--Schmidt changes the particular orthonormal coordinates but not the target subspace. Once the original targets are expressed through their coefficients $\bm c^{(A)}$, the recovered capacities are independent of that coordinate choice.

The construction is directly empirical. For sampled labeled segments $\{\bm z_t^{(n)},y_{1t}^{(n)},\ldots,y_{Kt}^{(n)}\}_{n=1}^{N_s}$, use $\widehat\mu_A=N_s^{-1}\sum_n y_{At}^{(n)}$ and the sample inner product $\langle u,v\rangle_{N_s}=N_s^{-1}\sum_n u^{(n)}v^{(n)}$. Applying Eq.~\eqref{eq:app_gramschmidt} with these sample averages gives score values $\widehat s_A^{(n)}$ that are exactly centered and orthonormal under the empirical measure, up to numerical roundoff. The same labeled record therefore determines both the score directions and the coefficients of every original target in that basis; no model for $P_{\rm ref}(\bm z)$ is required.

\section{Lindblad jump operators for the driven spin reservoir}
\label{app:lindblad_jumps}

This appendix gives the jump-operator construction used to generate the two input-dependent channels $\Phi_\pm$ in the numerical model. During one input interval the transverse field $a$ is constant, so the system Hamiltonian is the time-independent operator
\begin{equation}
H(a)=\sum_{i=1}^{N}\left(h_i\sigma_i^z+a\sigma_i^x\right)
+\sum_{i<j}J_{ij}\sigma_i^x\sigma_j^x.
\label{eq:app_spin_hamiltonian}
\end{equation}
We take a weak system--bath interaction of the form
\begin{equation}
H_{\rm tot}=H(a)+H_B+\lambda S\otimes B,
\qquad
S=\sum_{i=1}^{N}\sigma_i^y,
\label{eq:app_total_hamiltonian}
\end{equation}
where $B$ acts only on the bath and the bath is in the thermal state $\rho_B\propto e^{-\beta H_B}$. The input changes $H(a)$ but not the physical system--bath coupling operator $S$. Consequently, the jump operators change with $a$ because the energy eigenbasis and Bohr frequencies of the reservoir change, not because a different bath coupling is imposed.

For a fixed value of $a$, write the spectral decomposition of the full interacting Hamiltonian as
\begin{equation}
H(a)=\sum_{\epsilon}\epsilon\,\Pi_{\epsilon}(a),
\qquad
\sum_{\epsilon}\Pi_{\epsilon}(a)=I,
\label{eq:app_spectral_decomposition}
\end{equation}
where $\Pi_\epsilon(a)$ projects onto the complete eigenspace of energy $\epsilon$. In the interaction picture generated by $H(a)$, the system coupling operator is
\begin{align}
S_a(t)
&=e^{iH(a)t}Se^{-iH(a)t}\nonumber\\
&=\sum_{\epsilon,\epsilon'}
 e^{-i(\epsilon'-\epsilon)t}
 \Pi_{\epsilon}(a)S\Pi_{\epsilon'}(a).
\label{eq:app_S_interaction_picture}
\end{align}
Grouping terms that oscillate with the same Bohr frequency
$\omega=\epsilon'-\epsilon$ gives
\begin{equation}
S_a(t)=\sum_{\omega}e^{-i\omega t}A_a(\omega),
\label{eq:app_S_frequency_decomposition}
\end{equation}
with
\begin{equation}
A_a(\omega)=
\sum_{\epsilon'-\epsilon=\omega}
\Pi_{\epsilon}(a)S\Pi_{\epsilon'}(a)
.
\label{eq:app_jump_operator}
\end{equation}
These frequency-resolved components are the global jump operators. Indeed,
\begin{align}
[H(a),A_a(\omega)]
&=\sum_{\epsilon'-\epsilon=\omega}
(\epsilon-\epsilon')\Pi_{\epsilon}S\Pi_{\epsilon'}\nonumber\\
&=-\omega A_a(\omega),
\label{eq:app_jump_eigenoperator}
\end{align}
so $A_a(\omega)$ is an eigenoperator of the commutator with eigenvalue $-\omega$. Because $S$ is Hermitian,
\begin{equation}
A_a(-\omega)=A_a^\dagger(\omega).
\label{eq:app_jump_adjoint}
\end{equation}
For a nondegenerate spectrum, Eq.~\eqref{eq:app_jump_operator} can be written more explicitly as
\begin{equation}
A_a(\omega)=
\sum_{\substack{m,n\\E_n(a)-E_m(a)=\omega}}
\langle m;a|S|n;a\rangle
|m;a\rangle\langle n;a|.
\label{eq:app_jump_non_degenerate}
\end{equation}
Thus a positive-frequency operator maps an eigenstate of energy $E_n$ to one of energy $E_m=E_n-\omega$: the reservoir loses energy $\omega$ to the bath. The projector form in Eq.~\eqref{eq:app_jump_operator} is the one used conceptually in the numerics because it also treats degeneracies correctly. In particular, the $\omega=0$ sector
\begin{equation}
A_a(0)=\sum_\epsilon \Pi_\epsilon(a)S\Pi_\epsilon(a)
\end{equation}
is retained; it contains the part of the bath coupling that acts within equal-energy subspaces.

Under the weak-coupling, Born--Markov, and full secular approximations, the reduced evolution during the interval is
\begin{equation}
\dot\rho=\mathcal L_a(\rho)
=-i[H(a),\rho]
+\sum_\omega \gamma(\omega)\,
\mathcal D[A_a(\omega)]\rho,
\label{eq:app_lindblad_generator}
\end{equation}
where the Lamb-shift term is omitted, as in the numerical implementation, and
\begin{equation}
\mathcal D[L]\rho=L\rho L^\dagger
-\frac12\{L^\dagger L,\rho\}.
\end{equation}
Microscopically, $\gamma(\omega)$ is the Fourier transform of the equilibrium bath correlation function $\langle B(t)B(0)\rangle$. Thermal equilibrium imposes the KMS relation
\begin{equation}
\gamma(-\omega)=e^{-\beta\omega}\gamma(\omega),
\qquad \omega>0,
\label{eq:app_kms}
\end{equation}
which fixes the ratio of excitation to relaxation rates and makes the Gibbs state of $H(a)$ stationary for a fixed input. Following Ref.~\cite{Cenedese2026}, we use a flat bath spectrum: all downward transitions have the same bare rate $\gamma$, while their upward partners are weighted by the Boltzmann factor in Eq.~\eqref{eq:app_kms}. The zero-frequency component is retained with the same flat-spectrum convention.

The CPTP map for one reservoir update is therefore
\begin{equation}
\Phi_a=e^{\Delta t\,\mathcal L_a},
\qquad
\Phi_\pm=\Phi_{a_0\pm\delta a}.
\label{eq:app_spin_channel}
\end{equation}
At each change of the binary input we keep the operator $S=\sum_i\sigma_i^y$ fixed, diagonalize the new many-body Hamiltonian $H(a_\pm)$, rebuild the frequency sectors $A_{a_\pm}(\omega)$ from Eq.~\eqref{eq:app_jump_operator}, and propagate for the interval $\Delta t$ with the corresponding generator. This is the global thermal evolution used in both numerical parts of Sec.~\ref{sec:numerics}.

\bibliography{references}

\end{document}